\documentclass[11pt]{article}
\usepackage{fancyhdr}
\usepackage{isomath}
\usepackage{amsmath}
\usepackage{amsbsy}
\usepackage{amssymb}
\usepackage{amscd}
\usepackage{amsfonts}
\usepackage{array}
\usepackage{graphicx}
\usepackage{verbatim}
\usepackage{euscript}
\usepackage{alltt}
\usepackage{stmaryrd}
\usepackage{relsize}
\usepackage{enumitem}
\usepackage[lofdepth,lotdepth]{subfig}
\usepackage[numbers]{natbib}
\usepackage{caption}
\usepackage{float}
\usepackage{multirow}
\usepackage{epsfig}
\usepackage{color}
\usepackage{soul}
\usepackage{comment}

\usepackage[]{url}

\DeclareGraphicsExtensions{.eps,.pdf,.png}
\usepackage{bigints}
\usepackage{mwe}
\usepackage{algorithm}
\usepackage{algpseudocode}
\usepackage{relsize}
\usepackage[bottom]{footmisc}
\usepackage{amsmath}
\usepackage{amsbsy}
\usepackage{amssymb}
\usepackage{amscd}
\usepackage{amsfonts}

\newcommand{\bs}{\boldsymbol}

\newcommand{\bfb}{{\mathbold b}}

\newcommand{\bfn}{{\mathbold n}}

\newcommand{\bft}{{\mathbold t}}
\newcommand{\bfu}{{\mathbold u}}

\newcommand{\bfT}{{\mathbold T}}

\newcommand{\beq}{\begin{equation}}
\newcommand{\eeq}{\end{equation}}
\newcommand{\beqs}{\begin{eqnarray}}
\newcommand{\eeqs}{\end{eqnarray}}
\newcommand{\beql}{\begin{equation} \label}

\newcommand{\grad}{\mathop{\rm grad}\nolimits}

\newcommand{\curl}{\mathop{\rm curl}\nolimits}

\usepackage[margin=1in]{geometry}

\date{}
\begin{document}
\title{Emergent fault friction and supershear in a continuum model of geophysical rupture}

\author{Abhishek Arora\thanks{Department of Civil \& Environmental Engineering, Carnegie Mellon University, Pittsburgh, PA 15213.}
\thanks{
Department of Civil \& Environmental Engineering, Vanderbilt University, Nashville, TN 37235, email: abhishek.arora@vanderbilt.edu.} $\qquad$ Amit Acharya\thanks{Department of Civil \& Environmental Engineering, and Center for Nonlinear Analysis, Carnegie Mellon University, Pittsburgh, PA 15213, email: acharyaamit@cmu.edu.}}
\maketitle

\begin{abstract}
\noindent Important physical observations in rupture dynamics such as static fault friction, short-slip, self-healing, and supershear phenomenon in cracks are studied. A continuum model of rupture dynamics is developed using the field dislocation mechanics (FDM) theory. The energy density function in our model encodes accepted and simple physical facts related to rocks and granular materials under compression. We work within a 2-dimensional ansatz of FDM where the rupture front is allowed to move only in a horizontal fault layer sandwiched between elastic blocks. Damage via the degradation of elastic modulus is allowed to occur only in the fault layer, characterized by the amount of plastic slip. The theory dictates the evolution equation of the plastic shear strain to be a Hamilton-Jacobi (H-J) equation, resulting in the representation of a propagating rupture front. A Central-Upwind scheme is used to solve the H-J equation. The rupture propagation is fully coupled to elastodynamics in the whole domain, and our simulations recover static friction laws as emergent features of our continuum model, without putting in by hand any such discontinuous criteria in our model. Estimates of material parameters of cohesion and friction angle are deduced. Short-slip and slip-weakening (crack-like) behaviors are also reproduced as a function of the degree of damage behind the rupture front. The long-time behavior of a moving rupture front is probed, and it is deduced that the equilibrium profiles under no shear stress are not traveling wave profiles under non-zero shear load in our model. However, it is shown that a traveling wave structure is likely attained in the limit of long times. Finally, a crack-like damage front is driven by an initial impact loading, and it is observed in our numerical simulations that an upper bound to the crack speed is the dilatational wave speed of the material unless the material is put under pre-stressed conditions, in which case supersonic motion can be obtained. Without pre-stress, intersonic (supershear) is recovered under appropriate conditions.
\end{abstract}

\section{Introduction} \label{introduction_rupture}
Static friction is the physical phenomenon where the slipping between two blocks is restricted up to some applied shear force, with pressure being imposed upon the blocks. In the context of geomaterials, frictional behavior is interpreted in terms of stresses and not in terms of forces, as in conventional observations of friction. A rupture may be considered as the study of mechanics of two large blocks of material slipping with respect to each other over a thin region which is very small compared to the size of the blocks. In the context of a fault layer sandwiched between elastic blocks, thresholds corresponding to static friction may exist in terms of shear stresses, as one may expect motion of the front only after sufficient energy has been provided to induce damage in the undamaged region ahead of the rupture front. 

Conventionally, rupture dynamics has been studied as a problem of friction between two bodies. A crack is assumed to exist behind the rupture front, and the crack faces transmit shear and normal stresses, where the amount of shear stress transmitted is governed by an \textit{assumed} friction law and is assumed to be continuous across the crack surface. It has been historically suggested that dynamic rupture may be studied using the framework of a dislocation propagating in a slip plane \cite{brune1970tectonic,nabarro1987theory}. However, it is important to note some key differences between a classical dislocation and a crack-like model for rupture. For a dislocation in an elastic medium, there is a displacement discontinuity of fixed magnitude behind the dislocation line. The stresses transmitted across the slipped region are not limited in principle, but, have to be continuous, except at the core singularity. For a crack, any traction profile can be imposed as a Neumann boundary condition on both crack faces, and the displacement jump is not restricted, unlike a dislocation which has a jump of fixed magnitude behind the front. Moreover, the classical theory of dislocations and its solution techniques do not allow for incorporating damage in the wake of the dislocation line, which is essential for modeling a rupture using dislocation theory. 

Recently, a partial differential equation (PDE) based framework called field dislocation mechanics (FDM) was used to study dynamic rupture \cite{zhang2015single}. They could predict short-slip duration from their simulations by allowing for the appropriate amount of \textit{elastic damage}, i.e., degradation in elastic modulus, behind the front. Conventionally, short-slip duration is predicted using assumed slip velocity-weakening friction laws based on the physical phenomenon of self-healing, as suggested in \cite{heaton1990evidence}, and later explored in \cite{marone1991mechanics, dieterich1992earthquake, marone1998laboratory, rice2001new}. The velocity-weakening frictional behavior is also observed in laboratory shearing experiments in \cite{dieterich1996implications,scholz1998earthquakes}. In \cite{dieterich1994direct}, a microscopic insight into the evolution of friction over a characteristic sliding distance was attributed to the replacement of the contact population with a new contact population under sliding and the contact population age.  

In the work related to characterizing fault-friction, the dependence, of the predicted slip instability for assumed friction laws, upon state and rate variables like slip rate, slip distance, and time-dependent static friction was studied in \cite{ruina1983slip}. Along similar lines, theoretical and experimental work related to the frictional behavior of rocks at low normal stress was performed in \cite{ruina1986some}. Later, the processes that lead to weakening such as thermal pressurization of pore fluid in and near the fault core, and also flash heating at highly stressed frictional microcontacts during rapid slip, were explored in \cite{rice2006heating} for large crustal faults. Recently, in \cite{hulikal2018static} static and sliding contact between rough surfaces was studied, and the contact between two surfaces was modeled using discrete deformable elements attached to elastic blocks and these elements are elastic/viscoplastic springs.

\begin{figure}[ht!]
    \centering
    \includegraphics[scale=0.4]{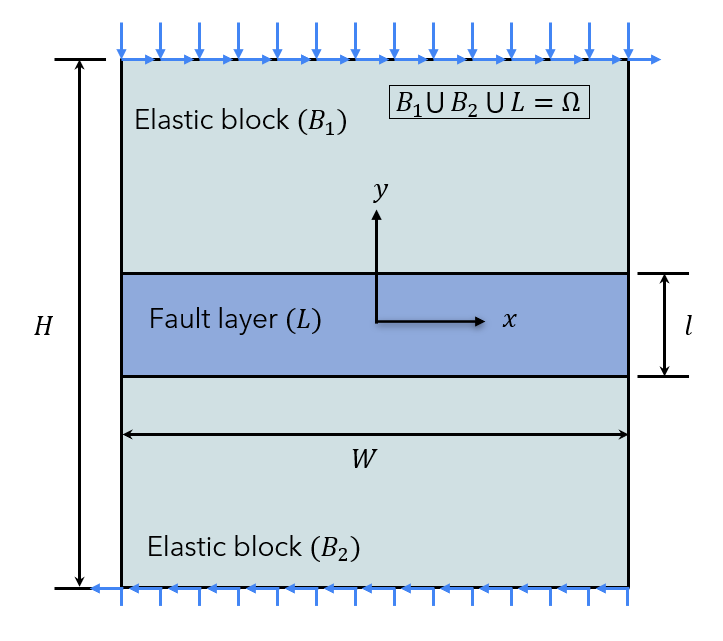}
    \caption{A schematic of a fault layer surrounded by elastic blocks and under transverse plus compressive loading.}
    \label{fig:layer_problem}
\end{figure}

In this work, we develop a continuum model for rupture, based on the rupture-related work in \cite{zhang2015single}. In their work, they utilized an ansatz to produce an exact, reduced, and planar model of FDM, where the dislocations or rupture front is allowed to move in a planar fault layer sandwiched between elastic blocks, as shown in Fig.~\ref{fig:layer_problem}. The amount of slip during rupture in the model is characterized by the plastic shear strain, and an evolution equation for the plastic shear strain can be derived by localizing the statement of the conservation of topological charge carried by a defect line, as derived for dislocations \cite{mura1963continuous,acharya2011microcanonical}, or for a crack-tip \cite{acharya2018fracture}. Finally, the evolution equation obtained for the  plastic shear strain is a Hamilton-Jacobi (H-J) equation and it enables the representation of a propagating rupture front. The plastic shear strain in the fault layer is assumed to induce elastic damage behind the rupture front, with the elastic blocks remaining undamaged. This leads to an energetic driving force contribution to the evolution of plastic strain, that was noted, but not accounted for, in the rupture-related simulations in \cite{zhang2015single}. 

We extend here the formulation of FDM \cite{zhang2015single} to model a propagating rupture front in the fault layer. We study fault friction and show that it emerges as a feature of our continuum model, although, there are no \textit{discontinuous} ``on-off'' criteria put in by hand in our model. 

\begin{figure}[ht!]
    \centering
    \subfloat[][$\eta = \eta_1 + \eta_2$]{
    \includegraphics[width=0.43\textwidth]{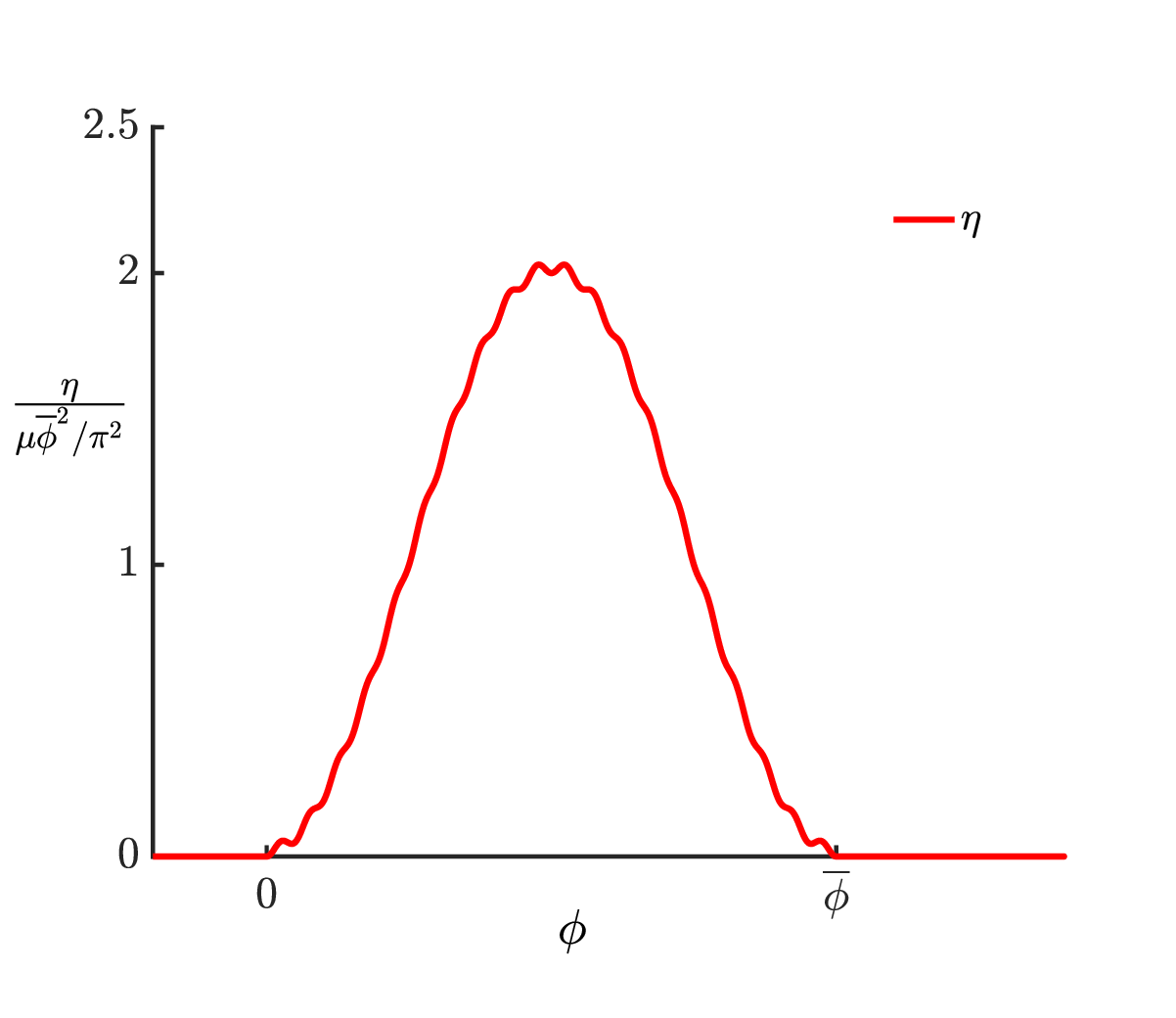}} \\
    \subfloat[][$\eta_1$]{
    \includegraphics[width=0.43\textwidth]{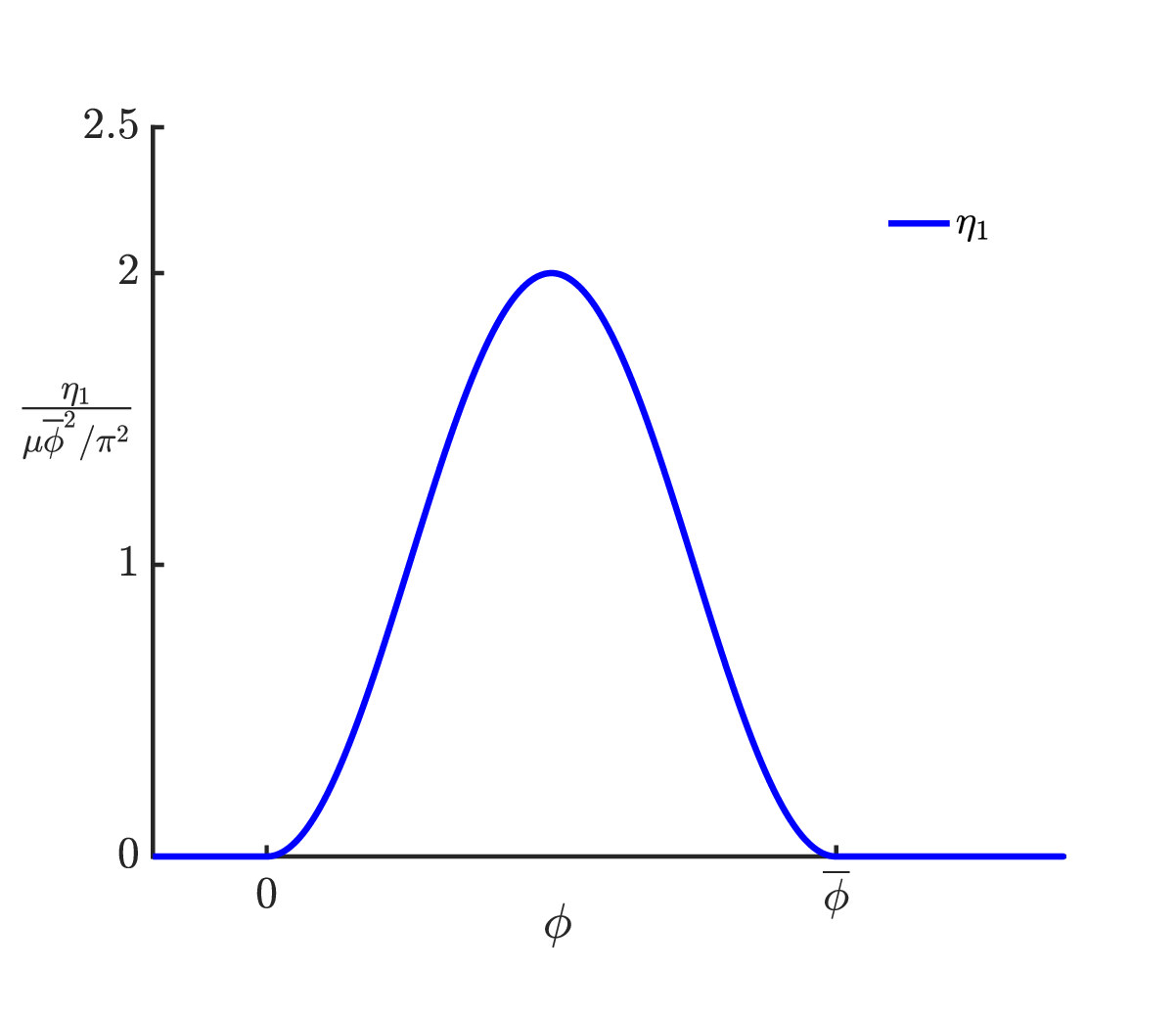}}
    \subfloat[][$\eta_2$]{
    \includegraphics[width=0.43\textwidth]{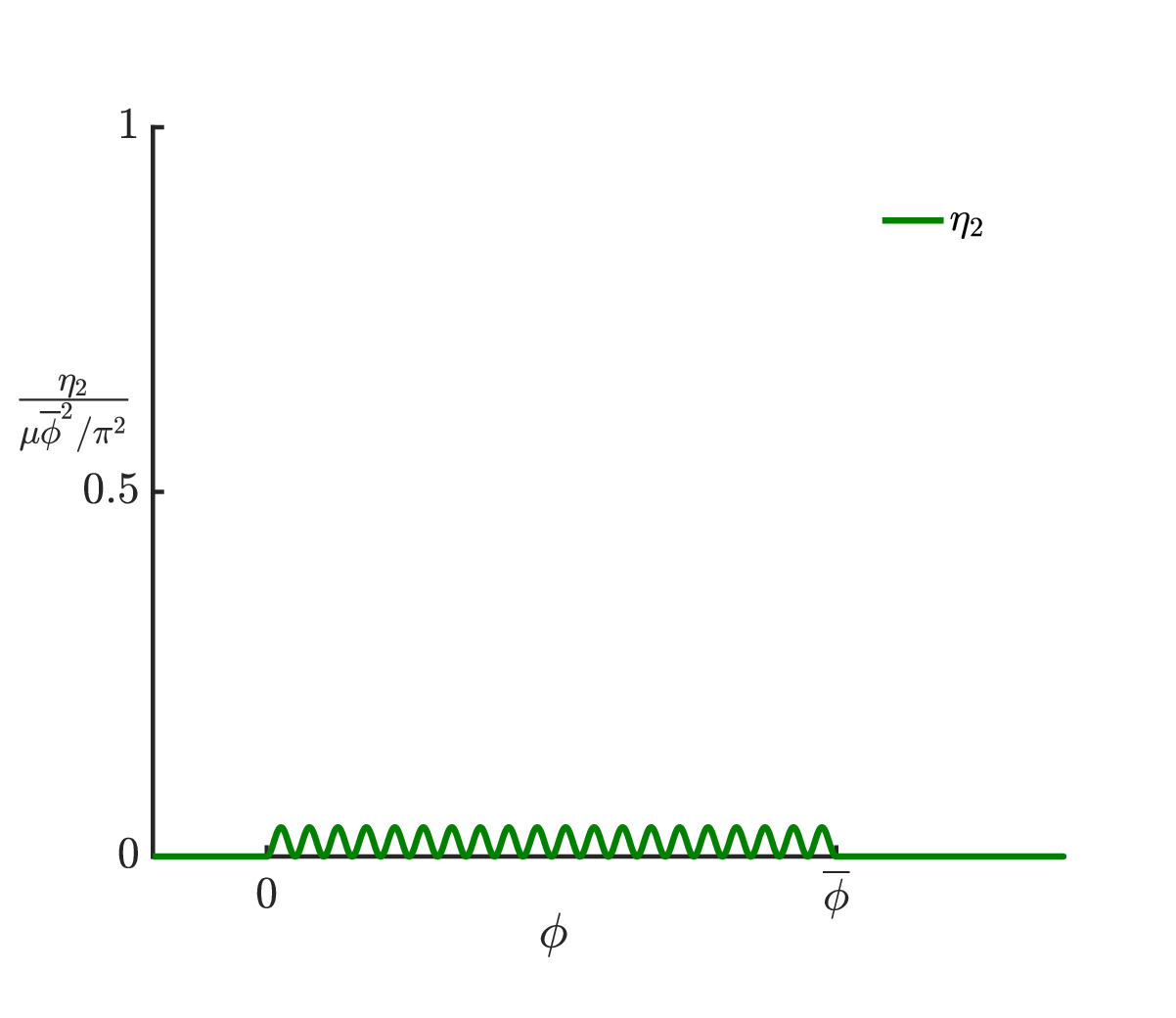}}
    \caption{A typical damage energy function in our model.}
    \label{fig:non-convex-function}
\end{figure}

\begin{figure}[htbp]
    \centering
    \includegraphics[scale=0.4]{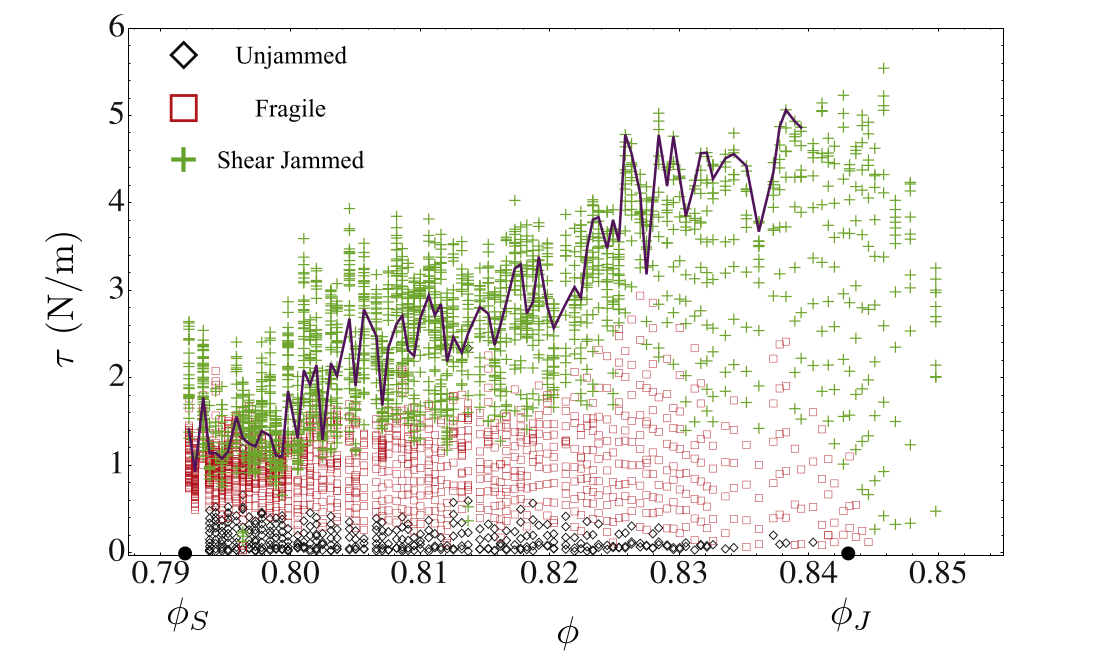}
    \caption{A compilation of measurements of shear stress ($\tau$) vs packing fraction $(\phi)$ (should not be confused with $\phi$ plastic strain in the current work) for granular materials, indicating whether the state is jammed, fragile, unjammed and at the yield stress. Figure reprinted from \cite{behringer2015jamming} with permission from \textit{Elsevier}.}
    \label{fig:jamming_plot_paper}
\end{figure}

The main physical assumptions built into our model are as follows: 
\begin{itemize}
    \item It is assumed that the fault layer responds to compressive elastic strain with intact compressive elastic stiffness, motivated by the observed behavior of geomaterials and the near impenetrability of crack flanks.
    \item There is an energetic cost required to induce elastic damage in the rupture front, which is modeled by a \textit{damage} energy density function $(\eta)$, shown in Fig.~\ref{fig:non-convex-function}. It is composed of two parts ($\eta_1$ and $\eta_2$), and $\eta_1$ models the fact that the damage energy required to cause motion increases only up to some value of slip/plastic strain. Hence, it monotonically increases, then decreases, and eventually becomes zero after a certain level of plastic strain is achieved. The $\eta_2$ function has small perturbations, with amplitude proportional to compressive elastic strain in the fault layer's normal direction, the latter reflecting the fact that geomaterials have higher strength in compression. The perturbations also vanish beyond the previously mentioned plastic strain/damage threshold (refer to Fig.~\ref{fig:non-convex-function}(c)). The perturbations in the $\eta$ function model the cyclic gain/loss in the strength of geomaterials due to rearrangement of their `fabric' \cite{cundall1979discrete, behringer2015jamming}. In \mbox{\cite[Fig.~11]{behringer2015jamming}}, which is reprinted here in Fig.~\ref{fig:jamming_plot_paper}, various states for granular materials are shown for different combinations of shear stress and packing fraction, like jammed, fragile, and unjammed states. At a given packing fraction, with an increase in the shear stress, an unjammed state transitions to a shear-jammed state and vice-versa, via the fragile state. In the context of our work, the perturbations in the damage-energy function model these relative states with lesser strength (unjammed), and shear-jammed states (higher strength). As the material gets sheared, it transitions between the relatively jammed and un-jammed states, which is reflected in the strength of the material. In addition to modeling this mechanical behavior, it is additionally assumed in our model that the relative increase/decrease in the strength of the material is proportional to the amplitude of compressive strain along the fault layer's normal direction, and it is absent in the case of tensile strains. 
    
   Our damage energy function is motivated by the wiggly energies model used for studying the kinetics of martensite transformation in \cite{abeyaratne1996kinetics} and using the same wiggly energy model, friction was studied in \cite{james1996wiggly} in a Hamiltonian setting, but had inconclusive results about the same. In \cite{mielke2012emergence}, a rate-independent dissipation (similar to dry friction) was deduced from a viscous system represented by wiggly energies, with a limit of vanishing viscosity. In the current work, as discussed previously, the oscillations in the damage energy function model the jammed and un-jammed states in geomaterials based on the experimental observations in \cite{behringer2015jamming}. These small and high-frequency perturbations in the energy result in a large energetic driving force for fault propagation, and subsequently have a significant impact on the fault dynamics and the prediction of pressure dependence of the strength against the rupture of the fault, as shown later in our numerical simulations.
    \item The rupture front velocity is assumed to be proportional to its theoretically derived driving force as shown in Sec.~\ref{theory_rupture}, a sufficient condition for the model to be in accord with the second law of thermodynamics. For a self-consistent evolution of a propagating rupture front, we account for the additional energetic driving force, mentioned earlier.
\end{itemize}  

In related work, a field crack mechanics model \cite{morin2021analysis} was developed and computed recently to model brittle crack propagation in elastic bodies. A key difference in that work from ours is that there is no internal stress due to a dislocation defect. However, the compressive elastic stiffness in the fault layer remains intact under compressive strain, as assumed here.

The outline of this paper is as follows: Sec.~\ref{theory_rupture} discusses the details of our mathematical model for a rupture front in the fault layer, while Sec.~\ref{numerical_scheme_rupture} outlines the numerical schemes used for solving the governing equations in our model. In Sec.~\ref{results_and_discussion_rupture}, we show that our numerical simulations recover the static friction laws as emergent features of our continuum model, also resembling the Mohr-Coulomb failure criterion for geomaterials, including well-defined features of cohesion and friction angle. We can also predict short-slip and self-healing features, which are physically observed during the occurrence of earthquakes, using the current model. Moreover, the long-time behavior of the moving rupture front is studied numerically, under applied shear and normal stresses. Finally, a body with a crack is put under impact loading conditions, and it is deduced that the front shows supershear behavior, i.e., the crack front moves at speeds higher than the shear wave speed of the bulk material. It is also shown that a limiting maximum speed of crack front propagation is the dilatational wave speed of the material when the boundary loading is solely an impulsively applied initial impact; supersonic motion of the crack front can be obtained under pre-strained conditions of a cracked body, as predicted previously in \cite{morin2021analysis, zhang2015single}.

\section{Theory}\label{theory_rupture}
The energy density function in our model has three ingredients i.e.~\textit{elastic}, \textit{core} and \textit{damage} energies, and is given by:
\begin{equation} \label{energy_density}
    \psi(\bs{\epsilon}^e, \bs{\alpha}, \bs{U}^p) + \eta(\bs{U}^p,\epsilon^e_{22}),
\end{equation}
where $\psi$ represents the summation of elastic and core energies, while $\eta$ is the damage energy, represented by a \textit{non-convex} function. We assume here that the elastic energy depends upon the plastic strain ($\bs{U}^p$) as well, in addition to elastic strain ($\bs{\epsilon}^e$), as the plastic straining induces damage which degrades the elastic moduli behind the rupture front. The core energy is an energetic penalty penalizing the sharpness of the rupture front, as the completely slipped or un-slipped fault layer is energetically preferred by the damage energy density. As discussed earlier, the damage energy density function is assumed to depend upon the plastic strain and the elastic compressive strain ($\epsilon^e_{22}$), as the strength of the material changes due to the re-arrangement of fabric, only under compressive strain. In \eqref{energy_density}, $\bs{\alpha}$ is Nye's dislocation density tensor which, for a given direction, yields the net Burgers vector content, per unit area, of dislocation lines threading the unit area perpendicular to the line direction.
In the fault layer, we utilize the ansatz from \cite{zhang2015single} and assume that the only non-zero plastic strain is:
\begin{subequations}
\begin{align}
    \bs{U}^p(x,y,t) &= U_{12}^p (x,t) \: \bs{e}_1 \otimes \bs{e}_2 := \phi(x,t) \: \bs{e}_1 \otimes \bs{e}_2, \nonumber 
\end{align}
\end{subequations}
and this physically affects the slip in the fault layer. Slip ($\delta$) is defined as the difference of the displacement along the horizontal direction $(u_1)$ between the top and bottom boundary of the fault layer. 

\noindent Using the incompatibility equation for small deformations, we get,
\begin{subequations}
\begin{align*}
    \bs{\alpha} (x,y,t) = - \curl \bs{U}^p (x,y,t) = - \phi_x(x,t) \: \bs{e}_1 \otimes \bs{e}_3, 
\end{align*}
\end{subequations}
and
\begin{equation*}
    \curl \bs{\alpha} =  \phi_{xx}(x,t) \: \bs{e}_1 \otimes \bs{e}_2.  
\end{equation*}
Further, we assume that the velocity of the rupture front in the fault layer is non-zero only along the $\bs{e}_1$ direction:
\begin{equation*}
\bs{V} (x,y,t) = V_1 (x,y,t) \bs{e_1} := v (x,t) \bs{e}_1.
\end{equation*}
With the above ansatz, the conservation law for the topological charge of a dislocation i.e., $\dot{\bs{\alpha}} = - \curl (\bs{\alpha} \times \bs{V})$ is simplified as shown below,
\begin{subequations}
\begin{align*}
    \phi_t(x,t) = - \phi_x v(x,t)
\end{align*}
\end{subequations}
(up to a gradient).

We assume linear elasticity and a quadratic form for the core energy:
\begin{equation} \label{eq:psi_assumption_rupture}
 \psi(\bs{\epsilon}^e, \bs{\alpha}, \phi) = \frac{1}{2} \bs{\epsilon}^e :  \mathbb{C} (\phi) : \bs{\epsilon}^e + \frac{1}{2}\epsilon |\bs{\alpha}|^2, 
\end{equation}
where $\epsilon$ is a parameter with physical dimensions of stress $\times$ length$^2$ and the plastic strain $(\phi)$ is used as an indicator of damage in the elastic modulus of the material in the fault layer. 
The elasticity tensor ($ \mathbb{C}$) is assumed to be of the following form:
\begin{equation*}
     \mathbb{C} := \left\{ 
    \begin{array}{l} 
    \bs{C}_L,
    \quad (\text{in} \:\: L) \\ 
    \bs{C}. \quad (\text{in} \:\: \Omega \backslash L)
   \end{array}
    \right\}
\end{equation*}
where $L$ is the fault layer, while $\Omega$ is defined as whole domain, as shown in Fig.~\ref{fig:layer_problem}.

\noindent Here, the tensor of elastic moduli ($\bs{C}$) in the elastic blocks ($\Omega \backslash L$) is assumed to be isotropic and can be written in indicial notation as shown below:
\begin{equation*}
    C_{ijkl} = \lambda \delta_{ij} \delta_{kl} + \mu(\delta_{ik} \delta_{jl} + \delta_{il} \delta_{jk}),
\end{equation*}
where $\lambda$ and $\mu$ are Lam\'e's constants. 

\noindent The elasticity tensor ($\bs{C}_L$) in the fault layer ($L$) is given as:
\begin{equation} \label{C_l_equation}
    \bs{C}_L = \left( \tilde{\lambda} - \frac{|\phi|}{\: \overline{\phi}} \kappa \right) \bs{C} + \left(1 - H(\epsilon^e_{22})\right) \frac{|\phi| \: }{\overline{\phi}} \: \kappa  \: C_{2222} \: \bs{e}_2 \otimes \bs{e}_2 \otimes \bs{e}_2 \otimes \bs{e}_2,
\end{equation} 
where, $ 0 < \kappa < \tilde{\lambda} < 1$. 

Here, $\tilde{\lambda}$ simply allows a difference in the elasticity of the intact fault material from the bulk elastic blocks, while $\kappa$ is a proportionality constant for damage induced in the ruptured regions of the fault layer due to the plastic straining. $H(.)$\footnote{$H(x)$ is defined as: $H(x) = 1$ if $x>0$, and $H(x) = 0$ if $x \leq 0$.} is the Heaviside step function, and the second term in \eqref{C_l_equation} allows the compressive elastic modulus in the fault to be intact under compressive strain. 

Now, as defined in \cite{zhang2015single}, the dissipation in our model is defined as the difference of power supplied by the external forces and the rate of change of free energy and kinetic energy in the whole domain, as shown below:
\begin{equation}
\begin{aligned}
    \mathcal{D} = \int_{\partial \Omega} \bft \cdot \dot{\bfu} \:  dl + \int_{\Omega} \bfb \cdot \dot{\bfu} \:  dA - \frac{d}{dt} \int_{\Omega}   \left( \psi(\bs{\epsilon}^e, \bs{\alpha}, \bs{U}^p) + \eta(\bs{U}^p,\epsilon^e_{22}) \right)\: dA 
    - \frac{d}{dt}\int_{\Omega} \left( \frac{1}{2} \rho |\dot{\bfu}|^2 \right) \: dA,
\end{aligned}
\end{equation}
where $\bft$ denotes the applied traction on the boundary of the whole domain, $\bfb$ is the applied body force in the whole domain, $\dot{\bfu}$ denotes the material velocity, and $\rho$ denotes the mass density.

The key difference between the current work and the rupture-related work in \cite{zhang2015single} is the inclusion of the driving force (for front propagation) due to the dependence of the elastic modulus (that will subsequently appear in the elastic strain energy density) on the damage characterized by the plastic strain. Other important extensions are an accounting of the wiggly perturbations in the damage energy density to model experimentally observed behavior in geomaterials \cite{behringer2015jamming}, and a dependence of the $\eta$ function on the compressive strain $\epsilon^e_{22}$ in the fault layer, as motivated previously. 

Now, assuming small deformation, and using the theorem of power expended, along with the additive decomposition of infinitesimal strain $(\bs{\epsilon})$ into elastic $(\bs{\epsilon}^e)$ and plastic $(\bs{U}^p)$ parts, one can obtain:
\begin{equation}
\begin{aligned}
     \mathcal{D} &=  \int_{L} \left( \bs{T} - \frac{\partial \psi}{\partial \bs{\epsilon}^e} - \frac{\partial \eta}{\partial \epsilon^e_{22}} \bs{e}_2 \otimes \bs{e}_2 \right) : \grad \dot{\bfu} \: dA + \int_{\Omega \backslash L} \left( \bs{T} - \frac{\partial \psi}{\partial \bs{\epsilon}^e} \right) : \grad \dot{\bfu} \: dA \\
     & +  \int_{L} \left( \frac{\partial \psi}{\partial \bs{\epsilon}^e} -  \frac{\partial \psi}{\partial \bs{U}^p} \right) : \dot{\bs{U}}^p \: dA  + \int_{L} \frac{\partial \psi}{\partial \bs{\alpha}} : \curl (\bs{\alpha} \times \bs{V}) \: dA   \\ &
      + \int_{L} \left(  \frac{\partial \eta}{\partial \epsilon^e_{22}} \bs{e}_2 \otimes \bs{e}_2 - \frac{\partial \eta}{\partial \bs{U}^p} \right): \dot{\bs{U}}^p \: dA.
\end{aligned}    
\end{equation}
In the absence of defects and plastic deformation, we obtain the constitutive relation for Cauchy stress as: 
\begin{equation}
    \bs{T} = \left\{
    \begin{aligned}
        & \frac{\partial \psi}{\partial \bs{\epsilon}^e} + \frac{\partial \eta}{\partial \epsilon^e_{22}} \bs{e}_2 \otimes \bs{e}_2, \quad (\text{in} \: L) \\
         & \frac{\partial \psi}{\partial \bs{\epsilon}^e} . \quad (\text{in} \:\: \Omega \backslash L).      
    \end{aligned}  
    \right\}
\end{equation}
Then, the expression for the dissipation becomes:
\begin{equation}
     \mathcal{D} =  \int_{L} \left( \bfT -   
     \frac{\partial \psi}{\partial \bs{U}^p} - \frac{\partial \eta}{\partial \bs{U}^p} \right) : \dot{\bs{U}}^p \: dA  + \int_{L} \frac{\partial \psi}{\partial \bs{\alpha}} : \curl (\bs{\alpha} \times \bs{V}) \: dA.
\end{equation}
Upon further manipulation, we get:
\begin{equation}
\begin{aligned}
     \mathcal{D} & = \int_{L} \left( \bfT -   
     \frac{\partial \psi}{\partial \bs{U}^p} - \frac{\partial \eta}{\partial \bs{U}^p} \right) : (\bs{\alpha} \times \bs{V}) \: dA  + \int_{L} \curl \left( \frac{\partial \psi}{\partial \bs{\alpha}} \right) :  (\bs{\alpha} \times \bs{V}) \: dA   \\ 
     & \quad \quad + \int_{\partial L} \frac{\partial \psi}{\partial \bs{\alpha}}: (\bs{\alpha} \times \bs{V}) \times \bfn \: dA
\end{aligned}    
\end{equation}
Using the assumptions of the ansatz as in \cite{zhang2015single}, the mechanical dissipation simplifies to:
\begin{equation*}
   \mathcal{D} = \int_L v(x,t) \left\{ 
         \left[ T_{12}(x,y,t) - \frac{\partial \eta}{\partial \phi}(x,y,t) - \frac{\partial \psi}{\partial \phi}(x,y,t)  + \epsilon \phi_{xx} (x,t) \right] (-\phi_x (x,t))
    \right\} dA.
\end{equation*}
The boundary integral for the horizontal boundary of the layer with the elastic material in the dissipation vanishes automatically with the assumption of the ansatz. For the vertical boundaries, for simplicity, it is assumed that $\phi_x$ is zero at those boundaries, and hence, the vertical boundary term does not contribute there as well. There are driving forces due to damage energy, elastic energy, and core energy, and some are responsible for causing or preventing the motion of the front, under external loads.

We assume the following expression of fault velocity such that it is proportional to the thermodynamically consistent driving force as shown below:
\begin{subequations}
\begin{gather*}
    v(x,t) := \frac{-1}{B} 
     \left\{ 
         \left[ \tau(x,t) - \tau^b (x,t) - \tau^{\psi}(x,t) + \epsilon \phi_{xx} (x,t) \right] \phi_x (x,t) 
    \right\}, \\
    \tau(x,t) := \frac{1}{2b} \int_{-b}^{b} T_{12} (x,y,t) \:dy, \quad
    \tau^{b}(x,t) := \frac{1}{2b} \int_{-b}^{b} \frac{\partial \eta}{\partial \phi} (x,y,t) \:dy,  \\
    \quad \tau^{\psi}(x,t) := \frac{1}{2b} \int_{-b}^{b} \frac{\partial \psi}{\partial \phi} (x,y,t) \:dy,
\end{gather*}
\end{subequations}
where $\tau$, $\tau^b$ and $\tau^{\psi}$ are the averaged driving force along $\bs{e}_2$ direction in the layer, due to shear stress, damage energy, and elastic energy dependence on plastic strain, respectively. Moreover, $B$ is a non-negative drag coefficient that characterizes the energy dissipation by specifying how the rupture front velocity responds to the applied driving force, locally. 

Now, substituting the expression of front velocity in $\mathcal{D}$, we get:
\begin{equation}
\begin{aligned}
  \mathcal{D} = \int_{L} \frac{1}{B} 
     \left\{ 
         \left[ \tau(x,t) - \tau^b (x,t) - \tau^{\psi}(x,t) + \epsilon \phi_{xx} (x,t) \right] \phi_x (x,t)
    \right\}^2 dx dy + \mathcal{R}
\end{aligned}    
\end{equation}
where
\begin{equation}
\begin{aligned}
    \mathcal{R} = & \int_{x = -W/2}^{x = W/2} -v(x,t) \phi_x(x,t) \left\{ \int_{-b}^{b} \Bigg[ \Big( T_{12} (x,y,t) - \tau(x,t) \Big) - \Bigg. \right. \\ \qquad
    & \left. \Bigg. \left( \frac{\partial \eta}{\partial \phi} (x,y,t) - \tau^b(x,t) \right) - \left( \frac{\partial \psi}{\partial \phi} (x,y,t) - \tau^{\psi}(x,t) \right) \Bigg] dy \right\} \: dx.
\end{aligned}    
\end{equation}
Finally, we deduce that
\begin{equation*}
    \mathcal{D} \ge 0 \quad \& \quad \mathcal{R} = 0.
\end{equation*}
As motivated in Sec.~\ref{introduction_rupture}, the damage energy density function is assumed to be of the following form:
\begin{equation}\label{eta_function}
    \eta := \mu \frac{\overline{\phi}^2}{\pi^2} \left\{  
    \begin{array}{c}
    q \left( 1- \text{cos} \bigg( 2 \pi \frac{|\phi|}{\:\overline{\phi}} \bigg) \right) +  R (-\epsilon^e_{22}) \left( 1- \text{cos} \bigg( 2 a \pi \frac{|\phi|}{\:\overline{\phi}} \bigg) \right), \: (\text{for} \: 0 \le \phi \le \overline{\phi}) 
    \\ 
    0, \quad (\text{otherwise})
    \end{array}
    \right\}
\end{equation}
where, $q$ controls the magnitude of the energetic cost required to induce damage in the fault layer, while $a$ is the frequency of the small perturbations w.r.t. plastic strain/damage. The magnitude of these small perturbations increases with an increase in compressive strain, as modeled by the ramp function $R(.)$\footnote{$R(x)$ is defined as: $R(x) = x$ if $x>0$, and $R(x) = 0$ if $x \leq 0$.} in \eqref{eta_function}. Also, $\overline{\phi}$ is the level of plastic strain after which the energetic cost required to induce damage becomes zero. The first and second terms in the function $\eta$ shown in \eqref{eta_function} correspond to the single hump $(\eta_1)$ and the high-frequency perturbation $(\eta_2)$ functions, respectively, as shown in an earlier example in Fig.~\ref{fig:non-convex-function}. The parameters taken for the example shown in Fig.~\ref{fig:non-convex-function} are $q=1$, $a=20$, with a uniform compressive strain $\epsilon^e_{22}=-0.02$ in fault layer.

\noindent Using the expression for $\eta$ in \eqref{eta_function} and for $\bs{C}_L$ in \eqref{C_l_equation}, the expression for driving forces due to damage energy is obtained as:
\begin{equation*}
\begin{split}
    & \tau^{b}(x,t) = \frac{1}{2b} \int_{-b}^{b} \frac{\partial \eta}{\partial \phi} (x,y,t) \:dy, \quad \text{where} \\
    & \frac{\partial \eta}{\partial \phi}  =   \frac{2 \mu \overline{\phi}\: \text{sgn}(\phi)}{\pi} \left\{ q   \: \text{sin} \bigg( 2 \pi \frac{|\phi|}{\: \overline{\phi}} \bigg)  + a R(-\epsilon^e_{22}) \: \text{sin} \bigg( 2 a \pi \frac{|\phi|}{\: \overline{\phi}} \bigg) \right\}.
\end{split}
\end{equation*}
Similarly, using the expression of $\psi$ in \eqref{eq:psi_assumption_rupture}, the driving force due to the dependence of elastic energy density on the plastic strain is obtained as:
\begin{equation*}
\begin{split}
    & \tau^{\psi}(x,t) = \frac{1}{2b} \int_{-b}^{b} \frac{\partial \psi}{\partial \phi} (x,y,t) \:dy, \quad \text{where}\\
    & \frac{\partial \psi}{\partial \phi}  = -\frac{\kappa}{\: \overline{\phi}}\:\text{sgn}(\phi) \left\{ \left( \frac{1}{2} \bs{\epsilon}^e : \bs{C} : \bs{\epsilon}^e \right) - (1-H(\epsilon^e_{22})) \: C_{2222} \: {\epsilon^e_{22}}^2 \right\}.
\end{split}
\end{equation*}
The final set of governing equations in our model are:
\begin{subequations}
\begin{gather}
    \rho \frac{\partial^2 u_i}{\partial t^2} = \frac{\partial T_{ij}}{\partial x_j}, \quad (\text{in} \:\: \Omega) \\
    \phi_t(x,t) =  - \phi_x v(x,t),  \quad (\text{in} \:\: L) \\
    v(x,t)  := \frac{-1}{B}
     \left\{ 
        \left(\phi_x\right) \left[ \tau - \tau^b - \tau^{\psi}  + \epsilon \phi_{xx} \right] 
    \right\}.
\end{gather}
\end{subequations}
where
\begin{equation} \label{eq:stress_constitutive_relation_rupture}
    T_{ij} := \left\{
    \begin{array}{c}
        {(C_L)}_{ijkl} (u_{k,l} - \phi \: \delta_{k1} \delta_{l2}) -
    H(-\epsilon^e_{22}) \mu \frac{\overline{\phi}^2}{\pi^2} \left( 1- \text{cos} \bigg( 2 a \pi \frac{|\phi|}{\:\overline{\phi}} \bigg) \right) \delta_{i2} \delta_{j2}, \:  (\text{in} \: L) \\
         C_{ijkl} \: u_{k,l} . \quad (\text{in} \:\: \Omega \backslash L)
    \end{array} 
    \right\}
\end{equation}
and $\bfu$ is the displacement vector.

For quasistatic loading, i.e., rates of loading that are very slow compared to the time scales of elastic wave propagation in the domain and also slower than the time scales of rupture front propagation (which is governed by the drag coefficient $B$), the dimensional equations that are to be solved are as follows:
\begin{equation}
\begin{aligned}
    &0 = \frac{\partial T_{ij}}{\partial x_j}, \quad (\text{in} \:\: \Omega), \\ 
    &0 =  \left(\phi_x\right)^2 \left[ \tau - \tau^b - \tau^{\psi}  + \epsilon \phi_{xx} \right]  \quad (\text{in} \:\: L).
\end{aligned}
\end{equation}
Taking $b$ as the fault zone width, and introducing the following non-dimensional variables for equations corresponding to $\textit{quasistatic evolution}$:
\begin{equation*}
    \Tilde{x} = \frac{x}{b}, \: 
    \Tilde{\bfu} = \frac{\bfu}{b}, \:
    \Tilde{\bs{T}} = \frac{\bs{T}}{\mu}, \:
    \Tilde{\tau}^{\psi} = \frac{\tau^{\psi}}{\mu}, \:
    \Tilde{\tau}^b = \frac{\tau^b}{\mu}, \:
    \Tilde{\epsilon} = \frac{\epsilon}{\mu b^2},
\end{equation*}
we obtain the following non-dimensional equations:
\begin{subequations}
\begin{gather}
    0 = \frac{\partial \Tilde{T}_{ij}}{\partial \Tilde{x}_j}, \quad (\text{in} \:\: \Omega) \label{eq:quasi_lmb} \\
    0 =  \left( \frac{\partial \phi}{\partial \Tilde{x}}\right)^2 \left[ \Tilde{\tau} - \Tilde{\tau}^b - \Tilde{\tau}^{\psi}  + \Tilde{\epsilon} \phi_{\Tilde{x} \Tilde{x}} \right]  \quad (\text{in} \:\: L) \label{eq:quasi_rupture}.
\end{gather}
\end{subequations}
Similarly, to get the non-dimensional form of governing equations for loading rates comparable with the time scales of elastic wave propagation and the time scales of rupture front evolution, the following non-dimensional variables are introduced:
\begin{equation*}
    \Tilde{x} = \frac{x}{b}, \: 
    \Tilde{t} = \frac{v_s t}{b}, \:
    \Tilde{\bfu} = \frac{\bfu}{b}, \:
    \Tilde{\bs{T}} = \frac{\bs{T}}{\mu}, \:
    \Tilde{\tau}^{\psi} = \frac{\tau^{\psi}}{\mu}, \:
    \Tilde{\tau}^b = \frac{\tau^b}{\mu}, \:
    \Tilde{\epsilon} = \frac{\epsilon}{\mu b^2}, \:
    \Tilde{B} = \frac{B b}{\sqrt{\mu \rho}} = \frac{v_s}{\mu/(B \: b)},
\end{equation*}
where $v_s = \sqrt{\mu/\rho}$ is the shear wave speed of the material, and the non-dimensional drag number $\tilde{B}$ represents the ratio of the elastic wave speed of the material to an intrinsic velocity scale of the motion of the rupture front. The non-dimensional form of the governing equations corresponding to $\textit{dynamic evolution}$ is:
\begin{subequations}
\begin{gather}
    \frac{\partial^2 \Tilde{u}_i}{\partial \Tilde{t}^2} = \frac{\partial \Tilde{T}_{ij}}{\partial \Tilde{x}_j}, \quad (\text{in} \:\: \Omega) \label{eq:dyn_lmb}\\
    \phi_{\Tilde{t}} = \frac{1}{\Tilde{B}} \left(\phi_{\Tilde{x}}\right)^2 \left[ \Tilde{\tau} - \Tilde{\tau}^b - \Tilde{\tau}^{\psi}  + \Tilde{\epsilon} \phi_{ \Tilde{x} \Tilde{x}} \right]  \quad (\text{in} \:\: L). \label{eq:dyn_rupture}
\end{gather}
\end{subequations}
As we are interested in studying the conditions leading to the motion of the rupture front, the displacement of the center of the rupture front (denoted by $u_1^c$) is written as a function of various non-dimensional ratios in our model. For computational purposes, we calculate $u_1^c$ by the displacement of the point corresponding to the maximum value of $\phi_x$ profile in the fault layer. Dimensional analysis suggests the following dependence of $u_1^c$ on the non-dimensional parameter ratios of our model:
\begin{subequations}
\begin{gather*}
    \frac{u_1^c}{b} = f^{*}\left( q, a, \frac{\lambda}{\mu}, \kappa, \tilde{\lambda}, \frac{\sigma_{ap}}{\mu}, \frac{\tau_{ap}}{\mu}, \frac{\epsilon}{\mu b^2}, \frac{B b}{\sqrt{\mu \rho}} \right), 
\end{gather*}
\end{subequations}
where $\sigma_{ap}$ and $\tau_{ap}$ are the applied compressive and shear stress, respectively, $f^*(\cdot)$ is some function of arguments shown above. Here onwards, the non-dimensional form of the governing equations is solved, \textit{and the non-dimensional quantities are written without} $\Tilde{()}$ \textit{for simplicity}. 
\section{Computational approximation of the PDEs} \label{numerical_scheme_rupture}
The numerical scheme used for solving the governing equations in our model is a mixed finite element method (FEM) and finite volume method (FVM) based technique. FEM is used for solving the linear momentum balance equation shown in \eqref{eq:quasi_lmb} or \eqref{eq:dyn_lmb}, while an FVM-based scheme is used for solving the H-J equation shown in \eqref{eq:quasi_rupture} or \eqref{eq:dyn_rupture}. The numerical schemes for the linear moment balance equation and H-J equation (both for $\textit{quasistatic and dynamic evolution}$) are discussed in the following sections. 
\subsection{Numerical scheme for scalar H-J equation} \label{sec:central_upwind_HJ_solve}
For both the quasistatic evolution and dynamic evolution, the H-J equation is solved using the same scheme. However, the $\phi_t$ introduced in the quasi-static evolution corresponds to a fictitious time (does not represent any physical time as in dynamic evolution). In this work, we are interested in obtaining equilibrated profiles of the rupture front under given loading conditions, and we use equations corresponding to quasistatic evolution to obtain an equilibrated state. With the fictitious time step introduced in the case of quasistatic evolution, we let the system evolve for given initial conditions. The overall idea for this scheme is similar to \cite{zhang2015single}, with a primary technical difference being the use of Central Upwind schemes of \cite{kurganov2000new, kurganov2001semidiscrete} to compute the nonlinear wave propagation in the fault layer.

The evolution equation for $\phi$ is split into convection and diffusion parts, as shown below:
\begin{equation}
\begin{aligned}
    \phi_t &= {\color{red} \frac{(\phi_x)^2}{B} \left[ \tau (x,t) - \tau^b (x,t) - \tau^{\psi}(x,t) \right]} + {\color{blue} \frac{(\phi_x)^2}{B} \epsilon \phi_{xx}}, \\
    & = {\color{red} H(\phi_x)} + {\color{blue} \frac{(\phi_x)^2}{B} \epsilon \phi_{xx}}
\end{aligned}    
\end{equation}
where, the term in red represents convection, and the term in blue models degenerate diffusion (as the diffusion is essentially localized at the rupture front). The convection terms are integrated explicitly, while the diffusion terms are implicitly integrated using the Crank-Nicolson method (second-order accurate in both space and time), as shown: 
\begin{equation}
\begin{aligned}
    \phi^{t+\Delta t}_i = & {\color{red} {(\phi_c)}^{t+ \Delta t}_i} + 
   {\color{blue}\frac{\Delta t}{2} \frac{\epsilon}{B}  \bigg[ \left( \frac{\phi^{t}_{i+1} - \phi^{t}_{i-1}}{2 \Delta h_x} \right) \frac{\phi^{t+\Delta t}_{i+1} - 2\phi^{t+\Delta t}_{i} + \phi^{t+\Delta t}_{i-1}}{2( \Delta h_x)^2} \bigg. +} \\ 
    & {\color{blue} \qquad \bigg.  \left( \frac{\phi^{t}_{i+1} - \phi^{t}_{i-1}}{2 \Delta h_x} \right) \frac{\phi^{t}_{i+1} - 2\phi^{t}_{i} + \phi^{t}_{i-1}}{2( \Delta h_x)^2}  \bigg]},
\end{aligned}    
\end{equation}
where ${(\phi_c)}^{t+ \Delta t}_i$ is the evolved $\phi$ due to convection, and it is obtained using a second order Runge-Kutta central upwind scheme \cite{kurganov2001semidiscrete}. Once, ${(\phi_c)}^{t+ \Delta t}_i$ is obtained, it can be seen that a matrix inversion is required to solve for $(\phi^{t+\Delta t})$, in the above scheme. 
\subsubsection{Central upwind scheme for convection part of H-J equation}
The $(\phi_c)^{t+\Delta t}$ at all grid points is given by $(\phi_c)^{t+\Delta t} = \phi^{(2)}$, which is obtained as shown below:

\begin{equation}
\begin{aligned}
     &\phi^{(1)}  = \phi^{t} + \Delta t \: C[\phi^t], \\
     & \phi^{(2)}  = \frac{1}{2} \phi^{t} + \frac{1}{2} \left( \phi^{(1)} + \Delta t \: C[\phi^{(1)}] \right),
\end{aligned}    
\end{equation}
where, $C[\phi]$ is given by, at $x=x_i$, as shown below
\begin{equation}
\begin{aligned}
    C_i = \frac{1}{a^{+}_{i} - a^{-}_{i}} \bigg[ a^{-}_{i} H(\phi^{+}_{x}) -  a^{+}_{i} H(\phi^{-}_{x})\bigg] - \frac{a^{+}_{i} a^{-}_{i}}{a^{+}_{i} - a^{-}_{i}} \left( \phi^{+}_{x} - \phi^{-}_{x} \right).
\end{aligned}    
\end{equation}
Here, $\phi^{\pm}_x$ are the right and left derivatives at the point $x=x_i$ of the reconstruction $\Tilde{\phi}(\cdot, t)$ at any time t, and $a^+_i$ and $a^-_i$ are the maximal one-sided local speeds, as defined below:

\begin{equation}
\begin{aligned}
    a^+_i := \text{max}\left\{ H'(\phi^+_x), H'(\phi^-_x), 0 \right\}, \\
    a^-_i := \text{min}\left\{ H'(\phi^+_x), H'(\phi^-_x), 0 \right\},
\end{aligned}    
\end{equation}
where $H'(\cdot)$ is the derivative of $H(\cdot)$ with respect to $\phi_x$. 

For a second-order accurate scheme in space, we use the continuous piece-wise quadratic polynomial for reconstruction $\Tilde{\phi}(\cdot, t)$, at time t, as shown below:
\begin{equation}
\begin{aligned}
    \Tilde{\phi}(x,t) := \phi_i + \frac{(\Delta \phi)_{i + \frac{1}{2}}}{\Delta h_x} (x - x_i) + \frac{(\Delta \phi)'_{i + \frac{1}{2}}}{2 (\Delta h_x)^2} (x - x_i) (x - x_{i+1}),
\end{aligned}    
\end{equation}
with $x \in [x_i ,x_{i+1}]$. Similarly, the reconstruction for $x \in [x_{i-1},x_{i}]$ is shown below:
\begin{equation}
\begin{aligned}
    \Tilde{\phi}(x,t) := \phi_i + \frac{(\Delta \phi)_{i - \frac{1}{2}}}{\Delta h_x} (x - x_i) + \frac{(\Delta \phi)'_{i - \frac{1}{2}}}{2 (\Delta h_x)^2} (x - x_i) (x - x_{i-1})
\end{aligned}    
\end{equation}
Using these, the right and left derivatives obtained at $x=x_i$ are 
\begin{equation}
\begin{aligned}
    \phi^{\pm}_x = \frac{(\Delta \phi)_{i \pm \frac{1}{2}}}{\Delta h_x} \mp \frac{(\Delta \phi)'_{i \pm \frac{1}{2}}}{2 (\Delta h_x)^2} 
\end{aligned}    
\end{equation}
In the above equations,
\begin{equation}
    (\Delta \phi)^t_{i +\frac{1}{2}} = \phi^t_{i+1} - \phi^t_i, \quad (\Delta \phi)^t_{i - \frac{1}{2}} = \phi^t_{i} - \phi^t_{i-1},
\end{equation}
and
\begin{equation}
\begin{aligned}
    (\Delta \phi)'_{i + \frac{1}{2}} = \text{minmod}\left( \theta \left[ (\Delta \phi)^t_{i +\frac{3}{2}} - (\Delta \phi)^t_{i +\frac{1}{2}} \right], \frac{1}{2} \left[ (\Delta \phi)^t_{i +\frac{3}{2}} - (\Delta \phi)^t_{i - \frac{1}{2}} \right], \right. \\ 
    \left. \theta \left[ (\Delta \phi)^t_{i +\frac{1}{2}} - (\Delta \phi)^t_{i - \frac{1}{2}} \right] \right)
\end{aligned}
\end{equation}
where $\theta \in [1,2]$, and the multivariable minmod$(\cdot)$ function is defined as:
\begin{equation}
\text{minmod}(x_1, x_2, \cdots) := \left\{ \begin{array}{c c}
     \text{min}_j \{x_j\} & \quad \text{if} \: \: x_j >0 \: \: \forall j, \\
     \text{max}_j \{x_j\} & \quad \text{if} \: \: x_j < 0 \: \:  \forall j, \\
     0 & \quad \text{otherwise}.
\end{array}
\right.
\end{equation}
For all simulations shown in this work, we take $\theta = 2$, as it corresponds to less dissipative limiters \cite{kurganov2001semidiscrete}.
\subsection{Displacement solve scheme: Quasistatic} \label{sec:quasi_static_disp_solve}
The weak form for the quasi-static displacement solve is shown below:
\begin{equation}
\begin{aligned}
   \int_{\Omega} \frac{\partial T_{ij}}{\partial x_j} \: \delta u_i \: dA = 0  \implies \int_{\Omega} T_{ij} \: \frac{\partial \delta u_i}{\partial x_j} \: d A = \int_{\partial \Omega} t_i \: \delta u_i \: dl,
\end{aligned}    
\end{equation}
where $t_i = T_{ij} n_j$ denotes the components of the traction vector, and $n_j$ is the components of the normal unit vector on the boundary.

Now, we substitute for the expression of Cauchy stress tensor from \eqref{eq:stress_constitutive_relation_rupture}, and we get the weak form as shown below:
\begin{equation} \label{eq:quasi-static_weak_form_layer}
\begin{aligned}
    & \int_{L} (C_L)_{ijkl} \left( \frac{\partial u_k}{\partial x_l} -\phi \: \delta_{k1} \delta_{l2} \right) \frac{\partial \delta u_i}{\partial x_j} \: d A  + \int_{\Omega/L} \left( C_{ijkl} \frac{\partial u_k}{\partial x_l}  \right) \frac{\partial \delta u_i}{\partial x_j} \: d A = \\ & \int_{\partial \Omega} t_i \: \delta u_i \: dl \: + 
    \int_{L} \left( H(-\epsilon^e_{22}) \mu \frac{\overline{\phi}^2}{\pi^2} \left( 1- \text{cos}  \bigg( 2 a \pi \frac{|\phi|}{\:\overline{\phi}} \bigg) \right) \delta_{i2} \delta_{j2} \right) \frac{\partial \delta u_i}{\partial x_j} \: dA.
\end{aligned}    
\end{equation}
It must be noted that the weak form shown in \eqref{eq:quasi-static_weak_form_layer} is nonlinear, as it is a function of displacement or equivalently elastic strain $(\epsilon^e_{22})$ for a fixed $\phi$. The above weak form is discretized using bilinear shape functions, and using isoparametric formulation. The LHS terms in \eqref{eq:quasi-static_weak_form_layer} contribute to the stiffness matrix (except for the term involving $\phi$ which goes into the residual vector), while the RHS terms contribute to the force vector. These are calculated from the corresponding element-wise calculations to obtain the global stiffness matrix and residual vector, as usual in any FEM implementation.
\subsection{Displacement solve scheme: Dynamic} \label{sec:dynamic_disp_solve}
In the case of dynamic problems, we use the centered difference explicit time integration scheme to solve the linear momentum balance equation. Based on the values of the previous time step $(\bfu^t, \dot{\bfu}^t,\ddot{\bfu}^t)$, one can first obtain the $\bfu^{t+\Delta t}$ using the following update, as shown:
\begin{equation}\label{eq:newmark_disp}
   \bfu^{t + \Delta t} = \bfu^t + \Delta t \: \dot{\bfu}^t  + \frac{(\Delta t)^2}{2} \: \ddot{\bfu}^t.
\end{equation}

\noindent Then, based on the updated displacement $(\bfu^{t+\Delta t})$, one can solve for acceleration $(\ddot{\bfu}^{t+\Delta t})$. The weak form for the acceleration solve is as follows:
\begin{equation}
   \int_{\Omega} \ddot{u_i}^{t+\Delta t} \delta u_i \: dA = - \int_{\Omega} T_{ij} \: \frac{\partial \delta u_i}{\partial x_j} \: d A + \int_{\partial \Omega} t_i \: \delta u_i \: dl.
\end{equation}
Now, substituting for the Cauchy stress tensor from \eqref{eq:stress_constitutive_relation_rupture}, we further get:
\begin{equation}\label{eq:newmark_accl}
\begin{aligned}
    \int_{\Omega} \ddot{u_i}^{t+\Delta t} \delta u_i \: dA & = - \int_{L} (C_L)_{ijkl} \left( \frac{\partial u_k}{\partial x_l} - \phi \: \delta_{k1} \delta_{l2} \right) \frac{\partial \delta u_i}{\partial x_j} \: d A  - \int_{\Omega/L} C_{ijkl} \frac{\partial u_k}{\partial x_l} \frac{\partial \delta u_i}{\partial x_j} \: d A \\ 
    & + \int_{\partial \Omega} t_i \: \delta u_i \: dl 
     + \int_{L} \left( H(-\epsilon^e_{22}) \mu \frac{\overline{\phi}^2}{\pi^2} \left( 1- \text{cos}  \bigg( 2 a \pi \frac{|\phi|}{\:\overline{\phi}} \bigg) \right) \delta_{i2} \delta_{j2} \right) \frac{\partial \delta u_i}{\partial x_j} \: dA.
\end{aligned}    
\end{equation}
The above weak form is discretized using bilinear shape functions and using the isoparametric formulation. The LHS terms contribute to the mass matrix, while the RHS terms contribute to the internal and external force vectors. These are calculated from the corresponding element-wise calculations to obtain the global mass matrix and force vector, as usual in FEM implementation. To avoid matrix inversion, we lump the global mass matrix using the row-sum lumping technique.

Finally, one can update the velocity using the following relation:
\begin{equation}\label{eq:newmark_vel}
    \dot{\bfu}^{t+\Delta t} = \dot{\bfu}^t + \frac{\Delta t}{2} \left( \ddot{\bfu}^{t} + \ddot{\bfu}^{t+\Delta t}\right)
\end{equation}
An initial condition for $(\bfu, \dot{\bfu})$ at $t=0$ is provided as data. 

The displacement update shown in \eqref{eq:newmark_disp} requires the acceleration ($\ddot{\bfu}^{0}$) at $t=0$. The only way to obtain the same is to evaluate the $rhs$ of \eqref{eq:newmark_accl} based on the initial conditions for displacement $(\bfu^{0})$ and $\phi$ at $t=0$.
\subsection{Algorithm}
The grid points for the FVM scheme in the fault layer are taken at the centers of elements of the FEM mesh. 

Time increment $(\Delta t)$ constraint for the stable evolution of state variables for quasistatic evolution (although, it is fictitious for quasistatic problem) is based on the $cfl$ condition in \cite{kurganov2000new}, and a constraint due to a source term as shown in \cite{zhang2015single}. For dynamic evolution, there is an additional time constraint due to the propagation of elastic waves \cite{zhang2015single, morin2021analysis}. To summarise, the time increment $(\Delta t)$ for both the quasistatic and dynamic evolution are given by: 
\begin{equation} \label{eq:time_increment_rupture}
\begin{aligned}
 \Delta t = \text{min}({\Delta t}_{\text{cfl}}, {\Delta t}_{\text{source}}), \quad (\text{for quasistatic evolution}) \\
 \Delta t = \text{min}({\Delta t}_{\text{cfl}}, {\Delta t}_{\text{source}}, {\Delta t}_{\text{dynamic}}), \quad (\text{for dynamic evolution})
\end{aligned}
\end{equation}
where
\begin{subequations}
\begin{align*}
  &\Delta t_{\text{cfl}} = 0.125 \frac{\Delta h_x}{\text{max}(|H'(x_i)|)}, \quad
 \Delta t_{\text{source}} = 0.1 \frac{B}{\text{max}(|s(x_i)|)}, \nonumber \\
&\Delta t_{\text{dynamic}} = 0.5 \frac{\text{min}(\Delta h_x, \Delta h_y)}{\sqrt{\frac{\lambda}{\mu} + 2}}.  \nonumber
\end{align*}
\end{subequations}
Here, $x_i$ is $x$ coordinate of grid points for $\phi$ solve, $\Delta h_x$ and $\Delta h_y$ is the minimum spacing in $x$ and $y$ direction in the fault layer, respectively, and $H'(\cdot)$ and $s(\cdot)$ are defined as shown below:
\begin{equation}
\begin{aligned}
   H'(x_i) & = \frac{2}{B} |\phi_x (x_i)| \Big( \tau (x_i) - \tau^b (x_i) - \tau^{\psi} (x_i)\Big), \\
   s(x_i) & =  \frac{(\phi_x (x_i) )^2}{B} \left[\left( \frac{\partial \tau}{\partial \phi} (x_i) - \frac{\partial \tau^b}{\partial \phi} (x_i) - \frac{\partial \tau^{\psi}}{\partial \phi} (x_i)  \right) \right].
\end{aligned}    
\end{equation}

\noindent The derivative of the driving force due to the averaged shear stress in the fault layer with respect to the plastic strain ($\phi$) is given by:
\begin{equation}
\begin{aligned}
    & \frac{\partial \tau}{\partial \phi} = \frac{1}{2b} \int_{-b}^{b} \frac{\partial T_{12}}{\partial \phi} dy, \quad \text{where} \\
    & \frac{\partial T_{12}}{\partial \phi} = - 2\mu \: \kappa \frac{\text{sgn}(\phi)}{\overline{\phi}}  \: \epsilon^e_{12} -  \mu \left( \Tilde{\lambda} - \frac{|\phi|}{\overline{\phi}} \kappa \right).
\end{aligned}  
\end{equation}
Similarly, the derivative of the driving force due to the damage energy function with respect to $\phi$ is given by:
\begin{equation}
\begin{aligned}
     & \frac{\partial \tau^b}{\partial \phi} = \frac{1}{2b} \int_{-b}^{b} \frac{\partial^2 \eta}{\partial \phi^2} dy, \quad \text{where} \\
    & \frac{\partial^2 \eta}{\partial \phi^2} = 4 \mu \: (\text{sgn}(\phi))^2 \left\{ q \: \text{cos} \left( 2\pi \frac{|\phi|}{\overline{\phi}} \right) + a^2 R(-\: \epsilon^e_{22}) \: \text{cos}\left( 2 a \pi \frac{|\phi|}{\overline{\phi}} \right) \right\} .
\end{aligned}    
\end{equation}
Finally, the derivative of the driving force due to the dependence of elastic energy on the plastic strain, with respect to $\phi$ is:
\begin{equation}
\begin{aligned}
     & \frac{\partial \tau^{\psi}}{\partial \phi} = \frac{1}{2b} \int_{-b}^{b} \frac{\partial^2 \psi}{\partial \phi^2} dy, \quad \text{where} \\
    & \frac{\partial^2 \psi}{\partial \phi^2} =  \frac{\kappa}{\: \overline{\phi}} \left( 2 \mu \: \text{sgn}(\phi) \: \epsilon^e_{12} -  \frac{\partial \text{sgn}(\phi)}{\partial \phi}\: \left\{ \left( \frac{1}{2} \bs{\epsilon}^e : \bs{C} : \bs{\epsilon}^e \right) - (1-H(\epsilon^e_{22})) \: C_{2222} \: {\epsilon^e_{22}}^2 \right\} \right).
\end{aligned}    
\end{equation}
In the above expression, the derivative $\partial \text{sgn}(\phi)/\partial \phi$ is obtained using a smooth $\text{sgn}(\phi) \approx \text{tanh} (\phi/c)$, with $c = 1 \times 10^{-4}$.

The algorithm for the $\textit{quasistatic evolution}$ and $\textit{dynamic evolution}$ is shown in Table \ref{tab:algorithm_quasistatic_rupture} and \ref{tab:algorithm_dynamic_rupture}, respectively.

\begin{table}[htbp]
{\begin{algorithm}[H]
\caption*{\textbf{Algorithm for quasistatic evolution}} 
\begin{algorithmic}
\State \textbf{Initialization}: For a given initial profile of $\phi^{(0)}$, an initial condition for $\bfu^{(0)}$ is obtained by solving \eqref{eq:quasi-static_weak_form_layer} for just one iteration (as the displacement solve is non-linear for a fixed $\phi$), for a given loading conditions. Choose a value for $tol$ and Set $n=0$. 
\\\hrulefill
\State \textbf{$\boldsymbol{{n+1}^{th}}$ iteration}: 
\begin{enumerate}
    \item  Known fields at $n^{th}$ iteration: $\phi^{(n)}$, $\bs{u}^{(n)}$.
    \item $\phi^{(n+1)}$ is obtained as discussed in Sec.~\ref{sec:central_upwind_HJ_solve}, with the fictitious time increment $\Delta t$ taken as shown in \eqref{eq:time_increment_rupture}.
    \item $\bs{u}^{(n+1)}$ is obtained by solving \eqref{eq:quasi-static_weak_form_layer} for just single iteration. The stiffness matrix and residual vector in \eqref{eq:quasi-static_weak_form_layer} for $\bs{u}^{(n+1)}$ solve is formed using $\phi^{(n+1)}$ and $\bs{u}^{(n)}$.
\end{enumerate}
\hrulefill
\State Set $n=n+1$, and repeat Steps $1-3$ until $\textit{norm}(\dot{\phi}) < tol$.
\end{algorithmic}
\end{algorithm}}
\caption{Algorithm for $\textit{quasistatic evolution}$.}
\label{tab:algorithm_quasistatic_rupture}
\end{table}

\begin{table}[htbp]
{\begin{algorithm}[H]
\caption*{\textbf{Algorithm for dynamic evolution}} 
\begin{algorithmic}
\State \textbf{Initialization}: Given: an initial profile of $\phi^{(0)}$ and initial conditions for $\bfu^{(0)}, \dot{\bfu}^{(0)}$, along with loading conditions. Set $n=0$.
\\\hrulefill
\State \textbf{$\boldsymbol{{n+1}^{th}}$ time step}: 
\begin{enumerate}
    \item  Known fields at $n^{th}$ time step: $\phi^{(n)}$, $\bs{u}^{(n)}, \dot{\bfu}^{(n)}, \ddot{\bfu}^{(n)}$.
    \item $\phi^{(n+1)}$ is obtained as discussed in Sec.~\ref{sec:central_upwind_HJ_solve}, with the time increment $\Delta t$ taken as shown in \eqref{eq:time_increment_rupture}.
    \item $\bs{u}^{(n+1)}$ is updated using \eqref{eq:newmark_disp}.
    \item $\ddot{\bs{u}}^{(n+1)}$ is obtained using \eqref{eq:newmark_accl}, with lumped mass matrix. The external and internal force vector in \eqref{eq:newmark_accl} for $\ddot{\bs{u}}^{(n+1)}$ solve is formed using $\phi^{(n+1)}$ and $\bs{u}^{(n+1)}$.
    \item $\dot{\bs{u}}^{(n+1)}$ is updated using \eqref{eq:newmark_vel}.
\end{enumerate}
\hrulefill
\State Set $n=n+1$, and repeat Steps $1-5$ until desired.
\end{algorithmic}
\end{algorithm}}
\caption{Algorithm for $\textit{dynamic evolution}$.}
\label{tab:algorithm_dynamic_rupture}
\end{table}
\section{Results and discussion} \label{results_and_discussion_rupture}
We first study fault friction using our theoretical and computational framework for rupture dynamics. For a given initial profile of the rupture front, we apply different levels of compressive and shear stress at the boundary, and we show through our simulations that thresholds in shear stress are obtained for a given compressive stress, beyond which the rupture front starts to propagate in the fault layer. Then, we apply larger shear stress than the threshold obtained previously, for a given compressive stress, and study the evolution of slip and the averaged shear stress in the fault layer. Moreover, the long-time behavior of a moving rupture front under shear stress is probed, and it is deduced that equilibrated rupture profiles without shear stress are not traveling wave profiles when shear stress is applied. Finally, we perform another case study, where a rupture front is driven by the impact loading, without any pre-stress in the domain, and it is deduced from our simulations, that an upper bound of the crack front speed is the same as the dilatational wave speed of the material. We reproduce supershear behavior as observed in the experiments of \cite{rosakis1999cracks}. However, it is only when the front is present under pre-stressed conditions, that our model can reproduce supersonic behavior.

The non-dimensional domain size, fault layer width, and mesh size in the fault layer for our simulations, unless mentioned otherwise, are shown in Table \ref{tab:simulation_details}.
\begin{table}[htbp]
    \centering
    \begin{tabular}{|c|c c c c c|}
    \hline
   \textbf{Length}  & $H/b$ & $W/b$ & $l/b$ & $\Delta h_x/b$ &  $\Delta h_y/b$ \\ 
         \hline
   \textbf{Value} & 100 & 100 & 2 & 0.2 & 0.2  \\
         \hline
    \end{tabular}
    \caption{Domain size, fault-layer width, and mesh size used for simulations.}
    \label{tab:simulation_details}
\end{table}

The parameters used for simulations here are given in Table \ref{tab:parameters}, unless mentioned otherwise. The only parameter that may change for different simulations is $\kappa$, depending upon the amount of elastic damage allowed in the wake of the rupture front in the fault layer.
\begin{table}[htbp]
    \centering
    \begin{tabular}{|c|c c c c c c c|}
    \hline
  \textbf{Parameter}  & $q$ & $a$ & $\epsilon/(\mu b^2)$ & $\lambda/ \mu$ & $B b/\sqrt{\mu \rho}$ & $\tilde{\lambda}$ & $\overline{\phi}$\\ 
         \hline
   \textbf{Value} & 1 & 100 & 10 & 2.25 & 1.0 & 0.7 & 0.5 \\
         \hline
    \end{tabular}
    \caption{Parameters used for simulations.}
    \label{tab:parameters}
\end{table}
\subsection{Emergent static fault friction} \label{sec:fault_friction}
The protocol for static fault friction simulations, with a given set of input parameters, is as follows:
\begin{enumerate}
    \item Quasistatic equilibrium is obtained without putting any external loads and an equilibrated $\phi$ profile is obtained.
    \item The equilibrated $\phi$ profile from Step 1 is taken as the initial condition, and different levels of normal stresses are applied. Quasistatic equilibrium corresponding to different normal stresses is obtained.
    \item The equilibrated $\phi$ profile corresponding to each normal stress case from Step 2 is taken as the initial condition, and different levels of shear stresses are applied. The highest level of shear stress for which the equilibrium is obtained is taken as the potential threshold in shear stress, corresponding to a given normal stress. 
    \item The equilibrated $\phi$ profile corresponding to the highest level of shear stress for a given normal stress, and the corresponding displacement field are taken as initial fields for a dynamic evolution. The threshold of shear stress is chosen based on the criteria that the center of the $\phi$ profile does not move until the stress waves generated from the rupture front are reflected back to the front from the boundary of the domain. \label{step_last}
\end{enumerate}

\begin{figure}[ht!]
    \centering
    \subfloat[][]{
    \includegraphics[width=0.48\textwidth]{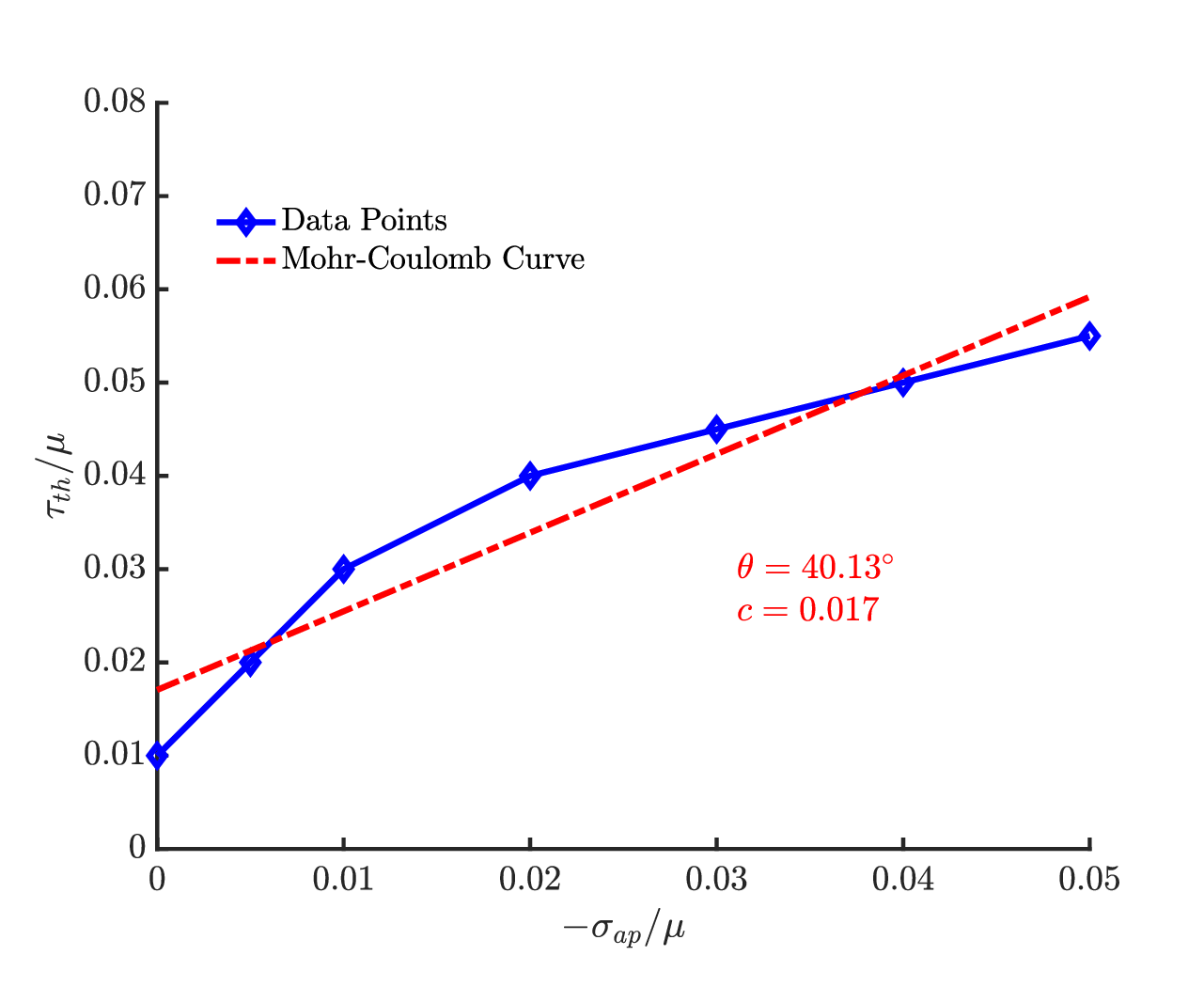}}
    \subfloat[][]{
    \includegraphics[width=0.48\textwidth]{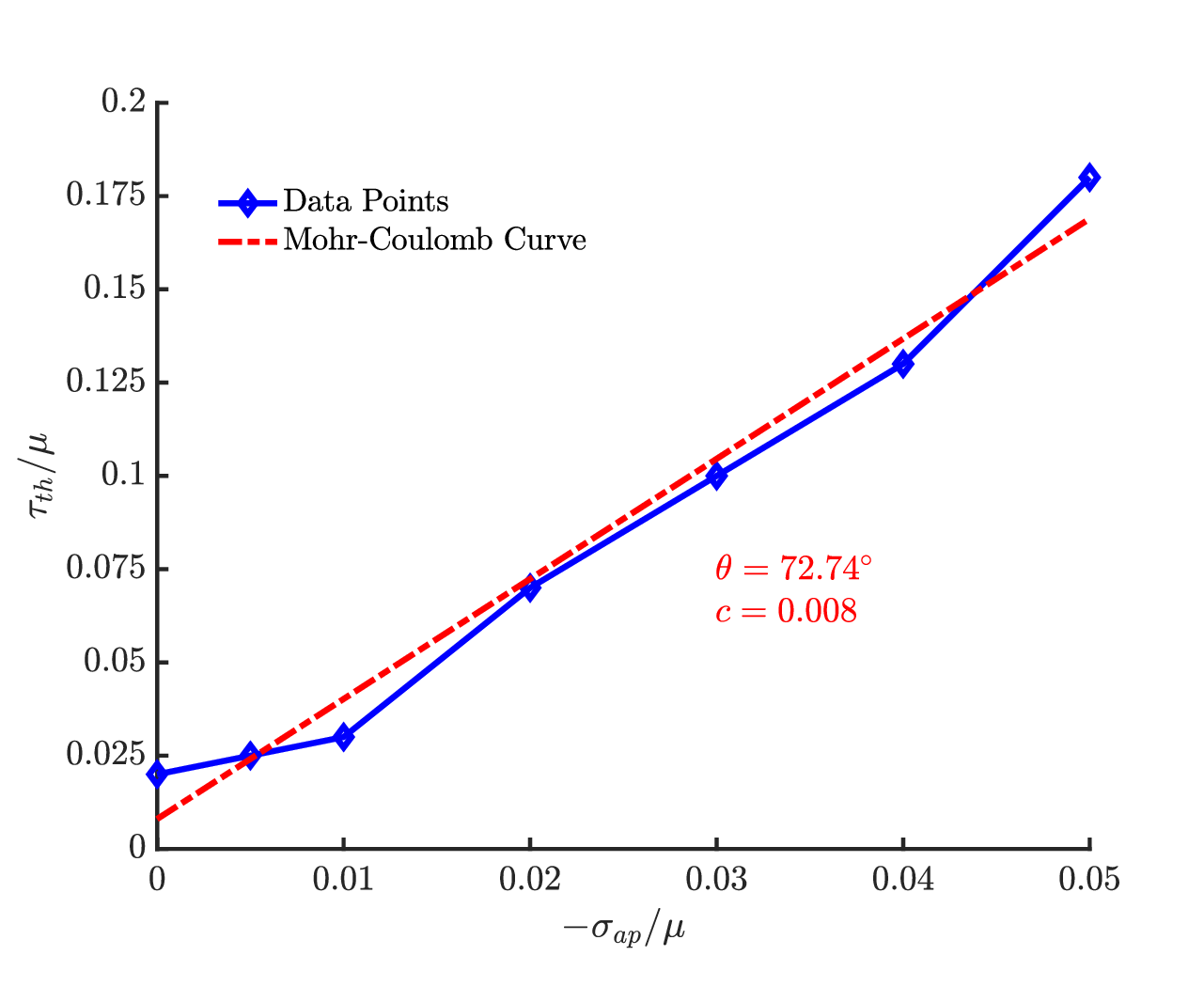}}
    \quad
    \subfloat[][]{
    \includegraphics[width=0.6\textwidth]{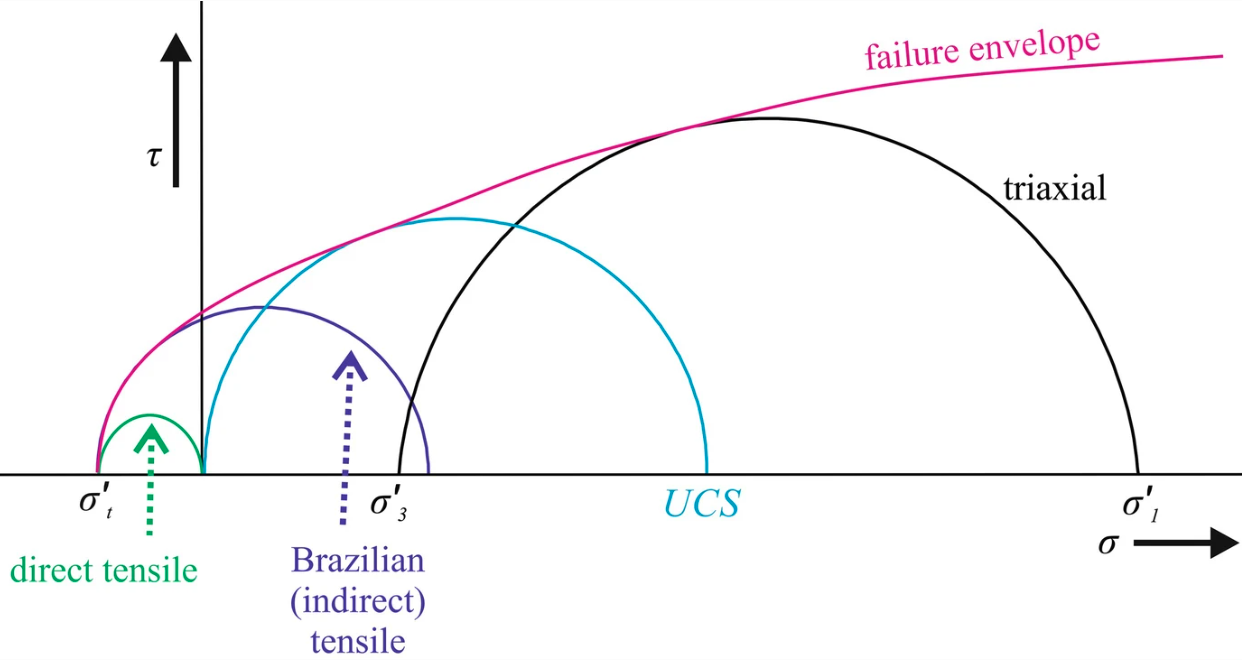}}
    \caption{(a) Shear stress thresholds ($\tau_{th}/\mu$) and the corresponding compressive stress ($-\sigma_{ap}/\mu$) obtained for the crack limit $(\kappa = 0.695)$, (b) for the lesser damage limit $(\kappa = 0.2)$. (c) The failure envelope for rocks based on experimental observations \cite{Hack2018}. Figure in (c) reprinted from \cite{Hack2018} with permission from \textit{Springer}.}
    \label{fig:friction_a_100_q_1_kappa_695}
\end{figure}

As discussed previously, a given state is considered to be equilibrated, if $norm(\dot{\phi}) < tol$, and the tolerance value $(tol)$ taken for the simulations reported in this work is $1.0 \times 10^{-3}$. It is important to note that we have not considered the convergence in the displacement field for judging the state of equilibria, as we only need an approximate force-equilibrated state for our purposes. This is also because our displacement solve is non-convex in $\epsilon^e_{22}$, due to the relative change in the strength of the fault layer material with the compressive strain, as modeled in the damage energy density function. However, in Step \ref{step_last}, it is observed that the rupture front does not move, even after accounting for the effects of inertia, for a given applied compressive stress and the corresponding threshold in shear stress. Moreover, to get confidence in the values of shear stress thresholds reported, the dynamic evolution is tested for the same equilibrated $\phi$ with a larger domain size, to make sure that the rupture front does not move in the time when stress waves are reflected back from the boundary to the front for a corresponding smaller domain size. 

In our model, there is competition between different energetic driving forces to prevent or enable motion, and this leads to thresholds for the movement of the rupture front, under given loading conditions. The driving force from the damage energy function prevents the motion of the front, while the averaged shear stress and strain energy terms enable the motion of the front. 

Based on the simulation protocol mentioned above, we obtain thresholds in shear stress for different normal stresses, and both the compressive stress and the shear stress are applied as traction boundary conditions. The computed envelope for the thresholds in shear stress versus the corresponding compressive stress from our simulations is shown in Fig.~\ref{fig:friction_a_100_q_1_kappa_695}, for both the crack limit $(\kappa = 0.695)$, and the lesser damage limit $(\kappa = 0.2)$. In both cases, the envelope obtained from our simulations resembles the textbook picture of a failure envelope for rocks which is based on experimental observations \cite{Hack2018}. The thresholds obtained for the same normal stress are higher for $\kappa =0.2$ as compared to that for $\kappa = 0.695$. This is due to the fact that there is an energetic advantage in moving the front toward the undamaged side, which is more prominent when the difference between the elastic modulus of the damaged and undamaged side of the front is more, i.e., for $\kappa = 0.695$. So, for a given compressive stress, the amount of shear stress required to move the front is more for $\kappa = 0.2$, as the front has a lesser amount of driving force to enable the motion of the front, as compared to that for $\kappa = 0.695$.

\begin{table}[ht!]
    \centering
    \begin{tabular}{|c|c c|}
    \hline
        \textbf{Damage parameter} $(\kappa)$ & Crack (0.695) & Lesser damage (0.2) \\ \hline
        \textbf{Cohesion} $(c)$ & $0.017 \: \mu$ & $0.008 \: \mu$ \\ 
        \textbf{Friction angle} $(\theta)$ & $40.13^{\circ}$ & $72.74^{\circ}$ \\ \hline
    \end{tabular}
    \caption{Prediction of material constants from our simulations for both the crack $(\kappa = 0.695)$ and the lesser damage limit $(\kappa = 0.2)$.}
    \label{tab:Prediction_for_constants}
\end{table}

Upon fitting the data points from our simulation results with a linear curve (that resembles a Mohr-Coulomb failure curve), we predict material constants for rocks. The cohesion value and the friction angle obtained for both limits are shown in Table \ref{tab:Prediction_for_constants}. The cohesion value is similar for both limits, but the friction angle obtained is higher for $\kappa =0.2$ as compared to that for $\kappa = 0.695$, as the thresholds are higher for a given applied compressive stress for $\kappa = 0.2$. The cohesion value reported in our simulations is correlated with the height of the damage energy function $(q)$ and the frequency of small perturbations in damage energy $(a)$, while the friction angle value is correlated with $a$.

\subsection{Short-slip duration and self-healing} \label{Sec:short_slip_and_self_healing}
We study the evolution of slip $(\delta)$ and the averaged shear stress $(\tau)$ in the fault layer, when a rupture front is driven due to the application of shear stress. This was studied in the rupture-related work in \cite{zhang2015single}, and physical observations of short-slip, slip-weakening, and self-healing were qualitatively reproduced in their numerical simulations. However, our model has more physical ingredients pertaining to geomaterials under multiaxial loading conditions. Hence, we probe similar physical phenomena using the theoretical framework developed in this work.

The non-dimensional domain size, fault layer width, and mesh size in the fault layer for the simulations reported in this Section are shown in Table \ref{tab:simulation_short_slip}. 
\begin{table}[ht!]
    \centering
    \begin{tabular}{|c|c c c c c|}
    \hline
    \textbf{Length}  & $H/b$ & $W/b$ & $l/b$ & $\Delta h_x/b$ &  $\Delta h_y/b$ \\ 
         \hline
    \textbf{Value} & 300 & 300 & 2 & 0.5 & 0.5  \\
         \hline
    \end{tabular}
    \caption{Domain size, fault-layer width, and mesh size used for short-slip simulations.}
    \label{tab:simulation_short_slip}
\end{table}

We consider $\text{dynamic evolution}$ when the rupture front is driven towards its undamaged side by applying higher levels of shear stress than the thresholds obtained for a given compressive stress for which the rupture front does not move, as shown in Sec.~\ref{sec:fault_friction}. A fixed observation point in the fault layer is chosen, and the slip, along with the averaged shear stress in the fault layer, is recorded, as the front moves towards the undamaged region of the fault layer. The protocol used for the simulations reported in this Section is as follows:
\begin{enumerate}
    \item The $\phi$ and $\bfu$ corresponding to the threshold value of shear stress for a given compressive stress, obtained as shown in Sec.~\ref{sec:fault_friction},  are taken as the initial conditions defining $\phi^{(0)}$ and $\bfu^{(0)}$, for the quasistatic solve discussed in Step 2.
    
    \item A quasistatic solve for a single iteration is done to obtain the additional displacement field, just due to the shear stress applied in addition to the threshold value of shear stress.
    
    \item The $\phi$ from Step 1 and the summation of $\bfu$ from Step 1 and Step 2 are taken as the initial conditions defining $\phi^{(0)}$ and $\bfu^{(0)}$, for the dynamic evolution runs. The initial condition for velocity is $\dot{\bfu}^{(0)} = \bf0$. The summation of the threshold value of shear stress and the additional shear stress applied in Step 2, along with the fixed compressive stress are applied as traction boundary conditions. Then, the system is allowed to evolve based on the governing equations of dynamic evolution.
\end{enumerate}

\begin{figure}[ht!]
    \centering
    \subfloat[][]{
    \includegraphics[width=0.45\textwidth]{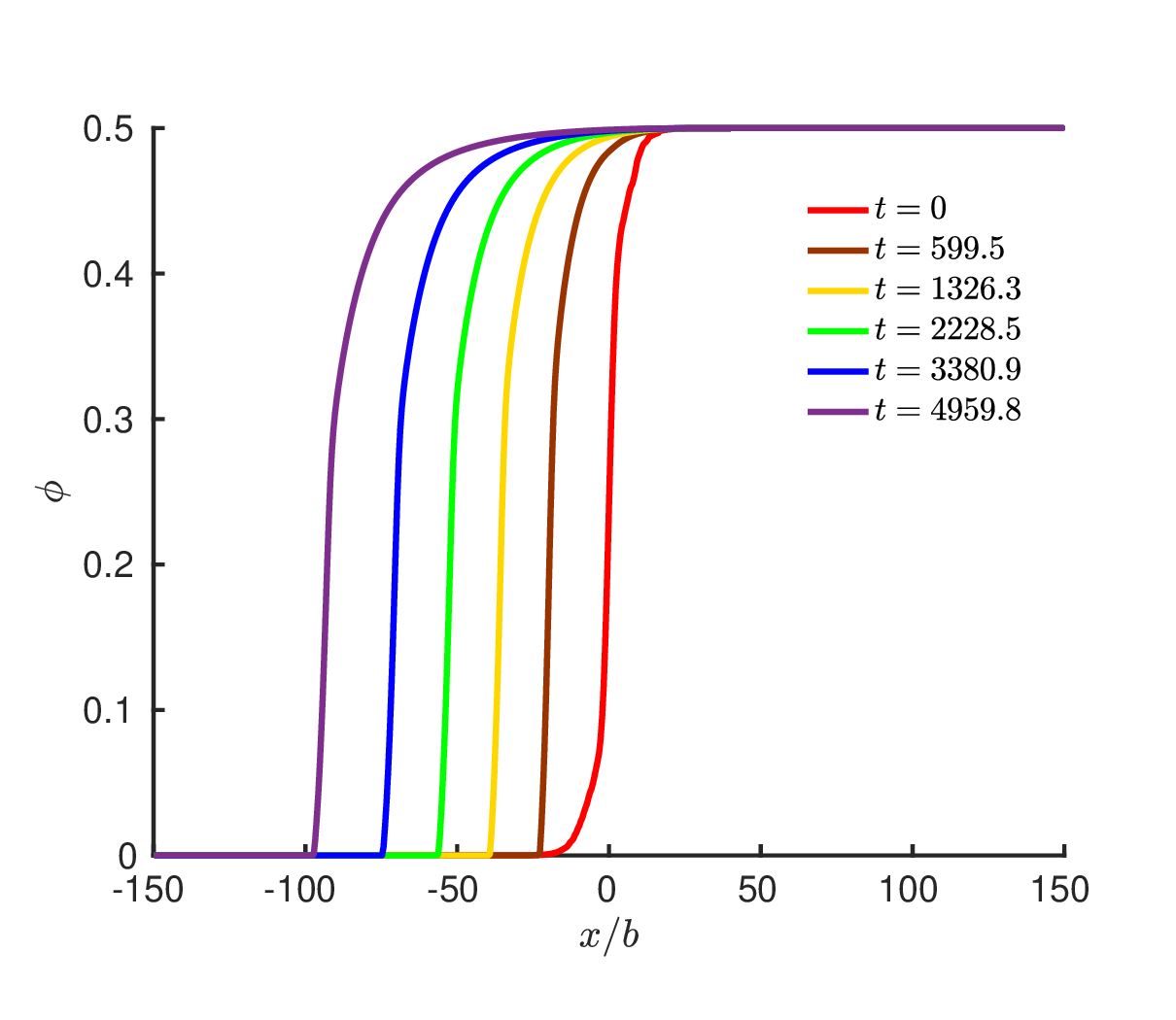}}
    \subfloat[][]{
    \includegraphics[width=0.45\textwidth]{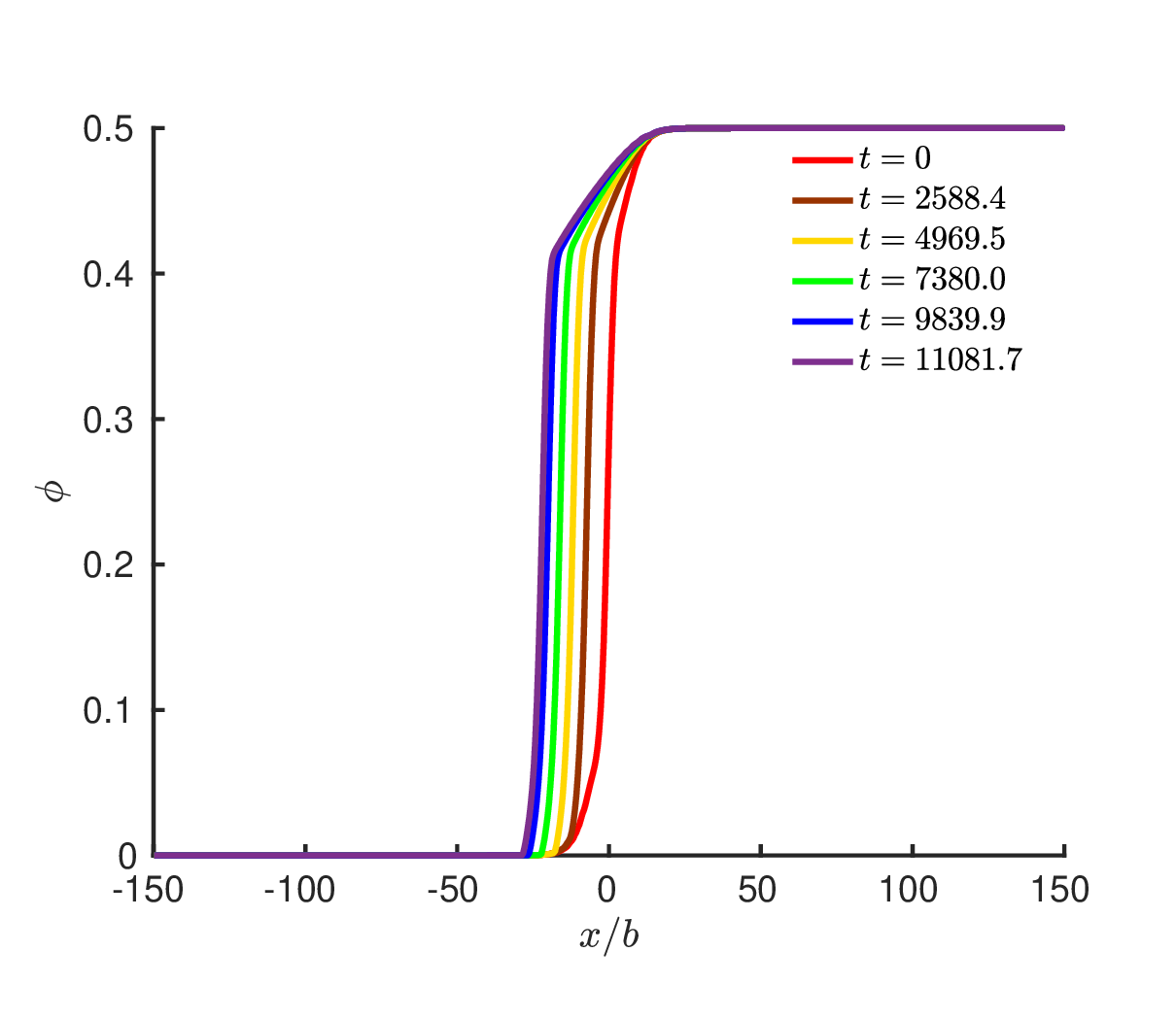}}
    \caption{The propagating rupture front under $\sigma_{ap}/ \mu = - 0.01$ and $\tau_{ap}/\mu = 0.1$ for (a) $\kappa = 0.695$, (b) $\kappa = 0.2$.}
    \label{fig:moving_front_rupture}
\end{figure}

\begin{figure}[ht!]
    \centering
    \includegraphics[scale=0.4]{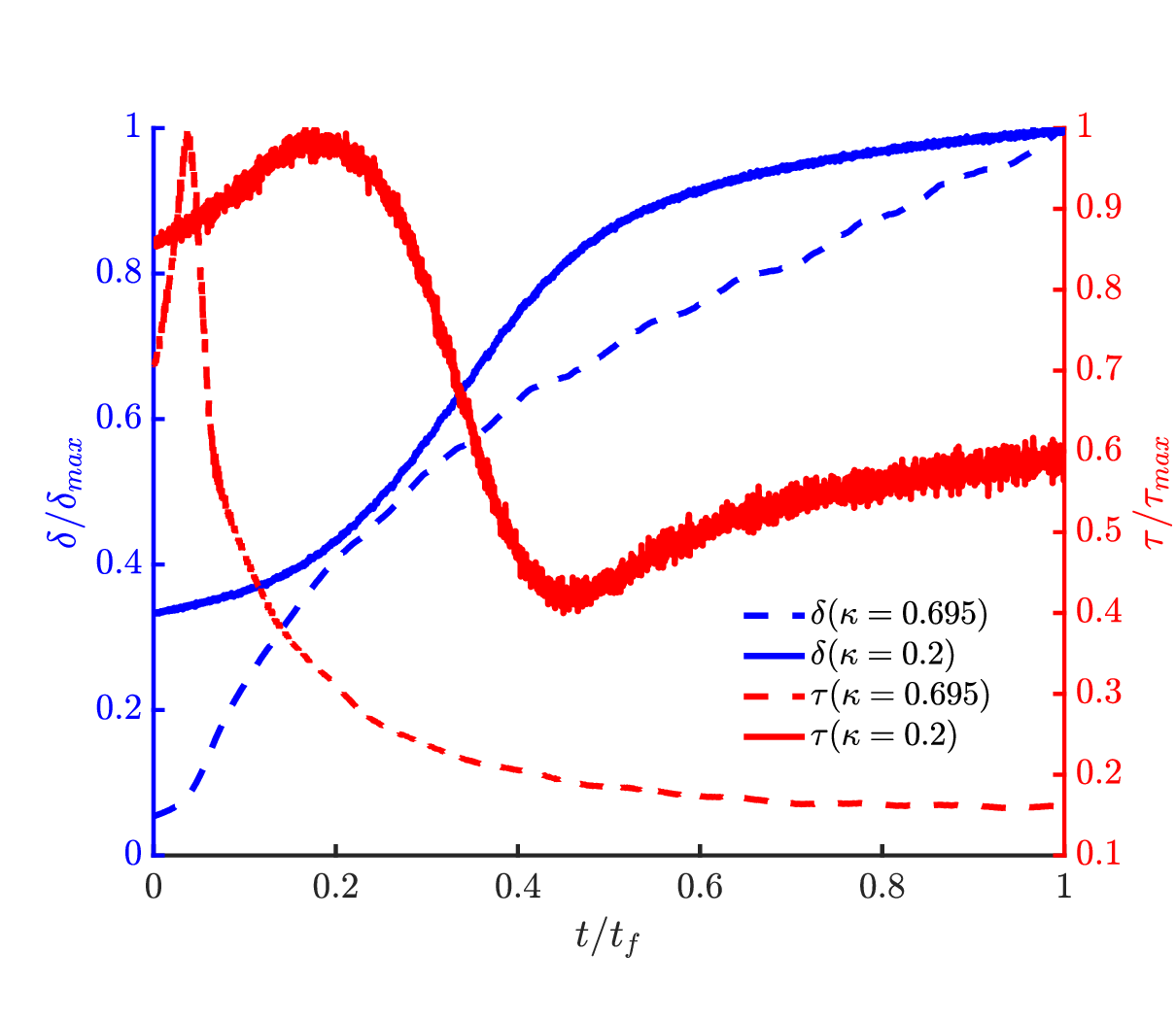}
    \caption{Short-slip and self-healing for the lesser damage limit $(\kappa = 0.2)$, and crack-like slip for the crack limit $(\kappa = 0.695)$.}
    \label{fig:short_slip}
\end{figure}

\begin{figure}[ht!]
    \centering
    \subfloat[][t=0]{
    \includegraphics[width=0.48\textwidth]{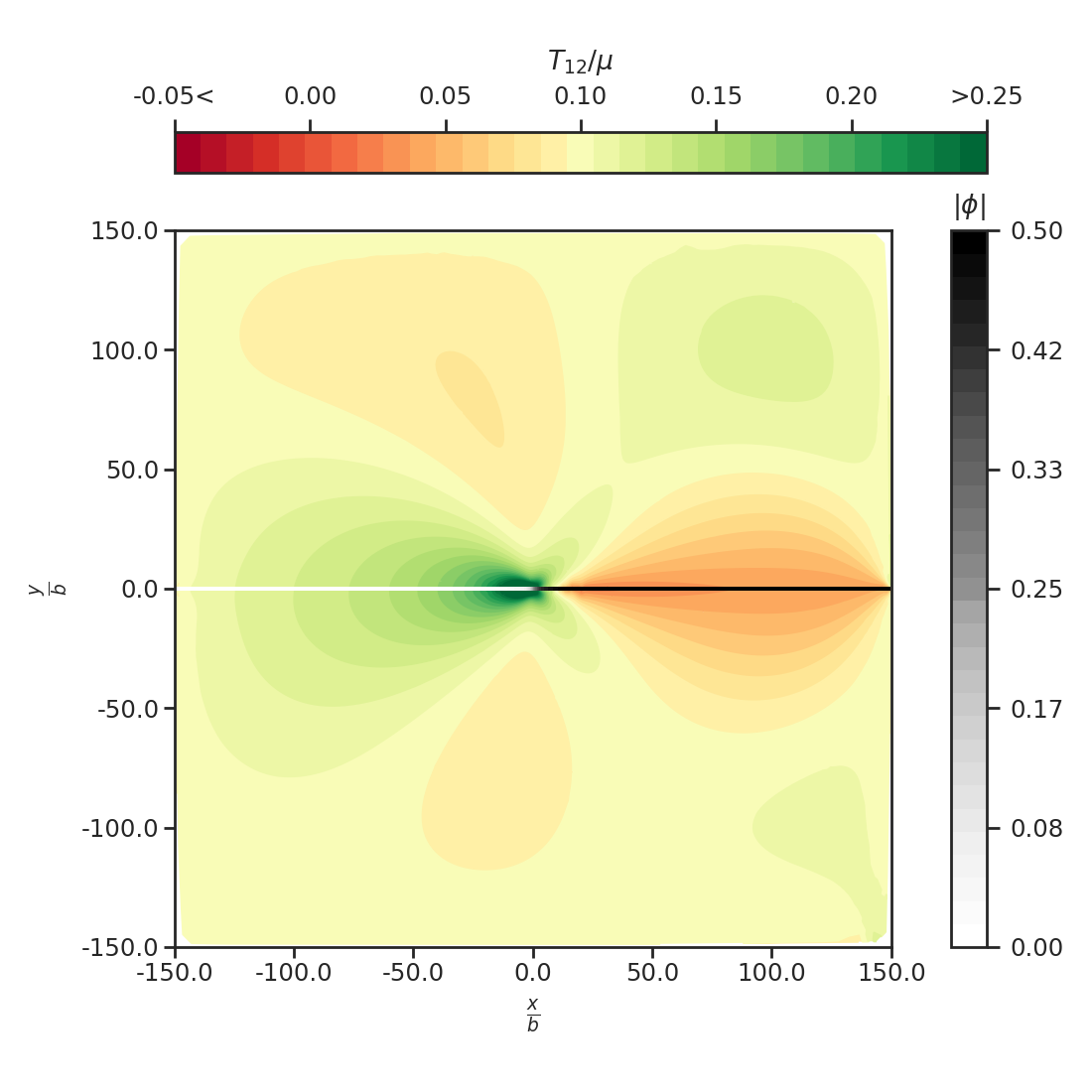}}
    \subfloat[][t=4959.88]{
    \includegraphics[width=0.48\textwidth]{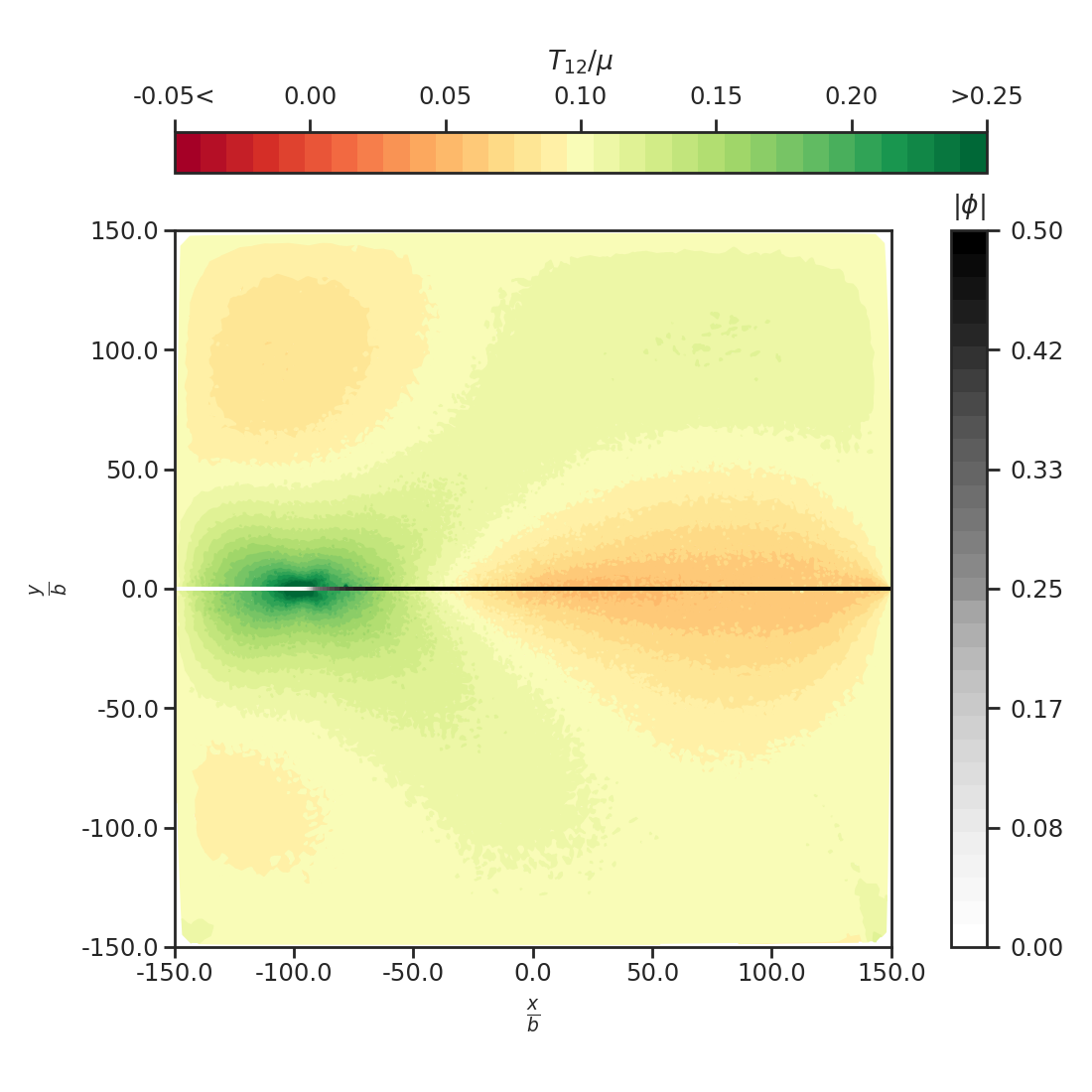}}
    \caption{The propagating rupture front for $\kappa= 0.695$ and the shear stress wave under $\tau_{ap}/ \mu = 0.1$ (pre-strain) and $\sigma_{ap}/ \mu = -0.01$, at different non-dimensional times.}
    \label{fig:subsonic_shear_wave_kappa_695}
\end{figure}

\begin{figure}[ht!]
    \centering
    \subfloat[][t=0]{
    \includegraphics[width=0.48\textwidth]{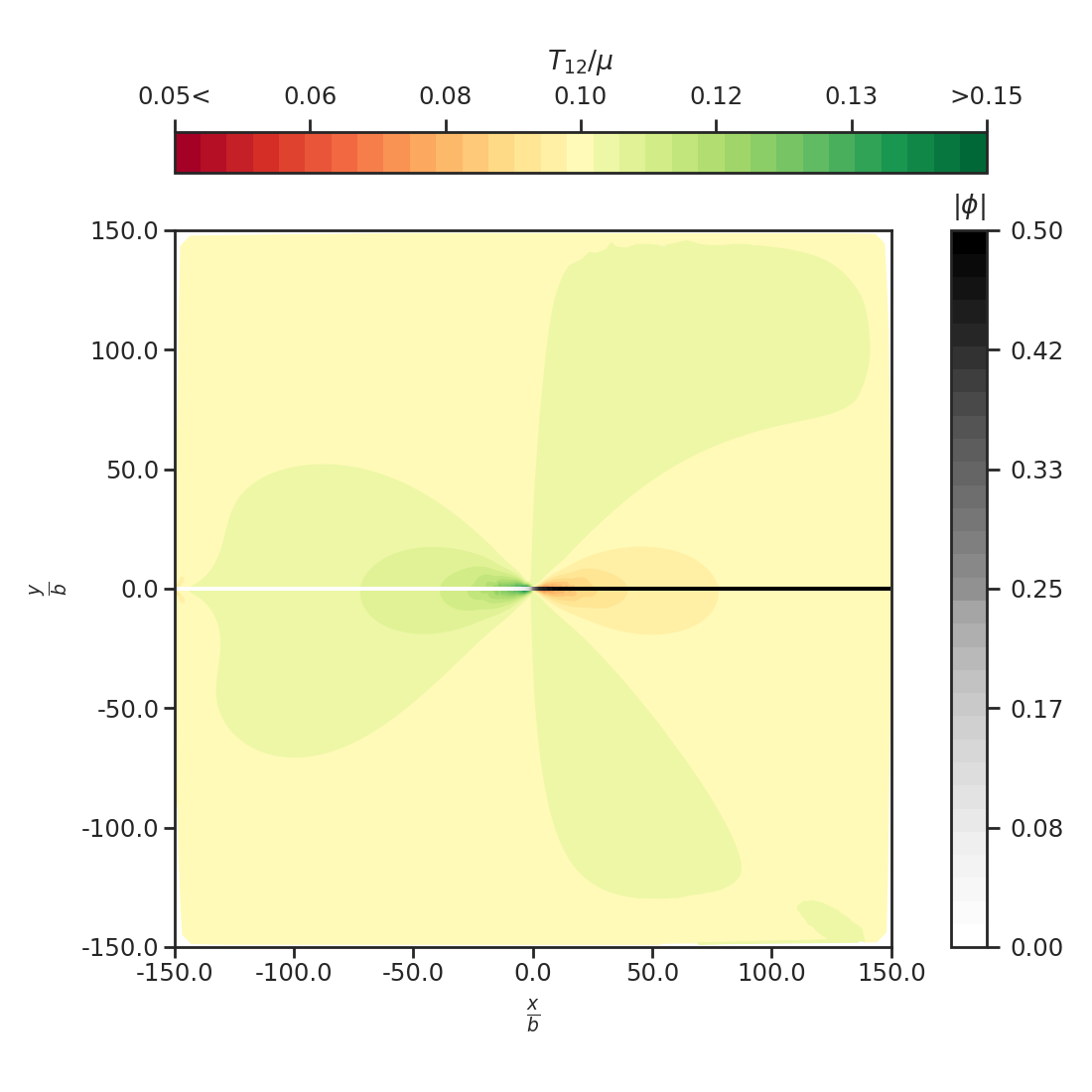}}
    \subfloat[][t=12328.85]{
    \includegraphics[width=0.48\textwidth]{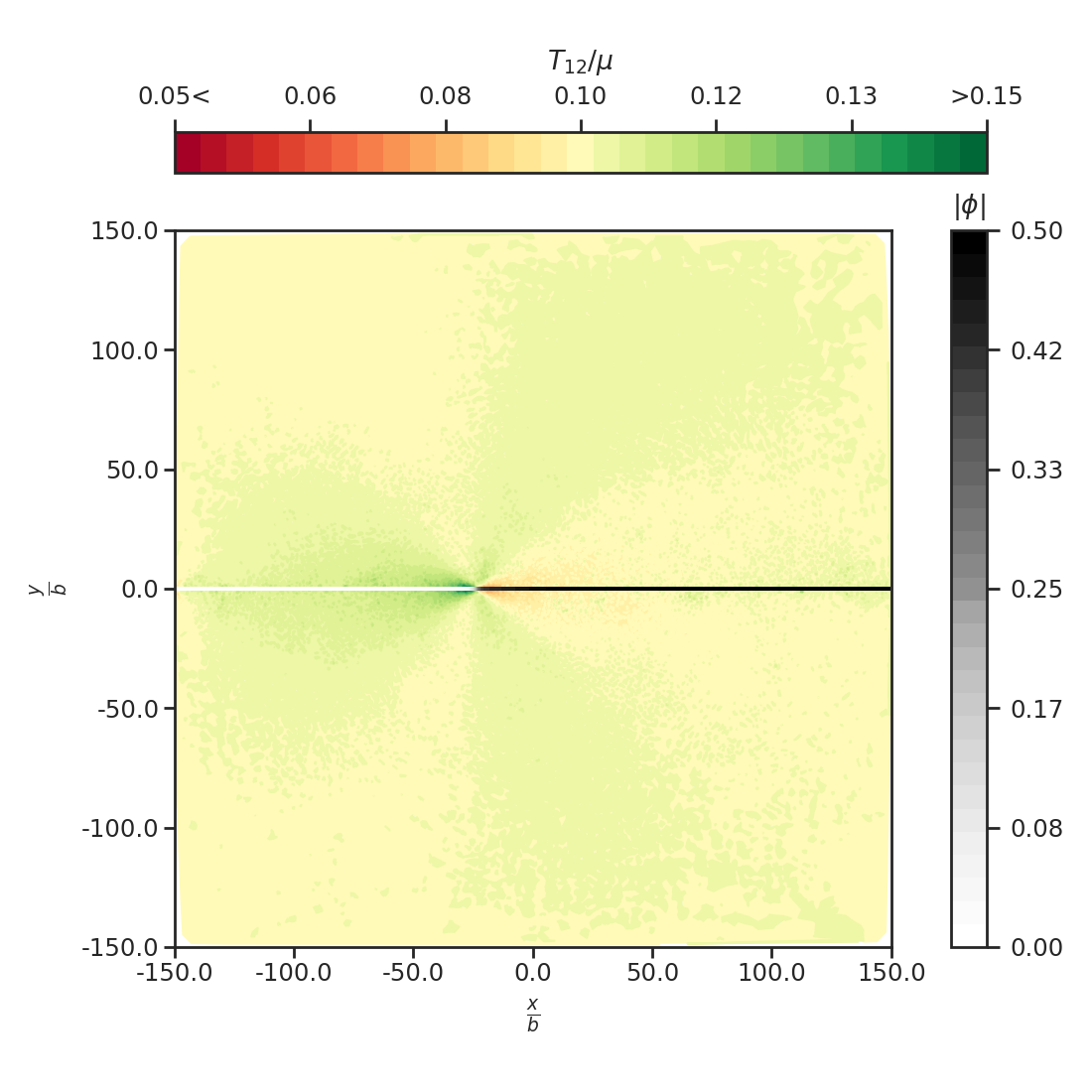}}
    \caption{The propagating rupture front for $\kappa= 0.2$ and the shear stress wave under $\tau_{ap}/ \mu = 0.1$ (pre-strain) and $\sigma_{ap}/ \mu = -0.01$, at different non-dimensional times.}
    \label{fig:subsonic_shear_wave_kappa_02}
\end{figure}

As discussed in the simulation protocol, the rupture front is driven under the normal stress ($\sigma_{ap}/ \mu = - 0.01 $), and the total shear stress  $(\tau_{ap}/\mu = 0.1)$, for both $\kappa = 0.2$ and $\kappa = 0.695$. Fig.~\ref{fig:moving_front_rupture} shows the propagation of the rupture front under the given applied stresses. For the same total applied shear stress, the amount of movement of the center of the rupture front is larger for $\kappa =0.695$, as compared to $\kappa = 0.2$, which is also consistent with the fact that the threshold of shear stress obtained for  $\kappa = 0.2$ was relatively larger.

The evolution of slip and the averaged shear stress at the observation point $(x/b = -10)$ in the fault layer is shown in Fig.~\ref{fig:short_slip}. The slip, the averaged shear stress, and the non-dimensional time in both cases are normalized with their maximum values, i.e., $\delta_{max}$, $\tau_{max}$, and $t_f$, respectively. It can be observed that the shear stress recovers in the lesser damage limit $(\kappa = 0.2)$, while it remains flat for the crack limit ($\kappa =0.695$). Similarly, it is also observed that the curve for slip vs time becomes flat for the lesser damage limit ($\kappa = 0.2$), while it continues to grow for the crack limit. Hence, it is evident that based on the amount of damage in the wake of the rupture front, our model reproduces different characteristics in slip and shear stress profiles. In particular, for the lesser damage limit, we can qualitatively predict short-slip and self-healing in geomaterials, which is observed physically during the occurrence of earthquakes \cite{heaton1990evidence}.  

We further plot the shear waves in the dynamic simulations in Step 3 for the crack case in   Fig.~\ref{fig:subsonic_shear_wave_kappa_695}, and for lesser damage case in Fig.~\ref{fig:subsonic_shear_wave_kappa_02}. It can be deduced from these Figures that there are no Mach cones in the corresponding shear waves, and the rupture front is moving subsonically in both cases. It is important to note that based on the simulation protocol used here, the rupture front is moving under pre-strain (as a result of Step 2 in the protocol).

\subsection{Long-time behavior of the moving rupture front}
We consider the crack case ($\kappa = 0.695$) of the rupture front for studying the long-time behavior of the front under external tractions. In dynamic evolution, it can be difficult to study such long-time behavior of the front without incurring the effects of reflected stress waves from the boundaries, as that would require doing such simulations on very large domains. An alternative is to do quasistatic evolution, at loading rates much slower than the time scales of elastic wave propagation and the time scales of rupture front motion. Quasistatic evolution also has a limitation where the domain shape and size play a role in the stress fields produced in the domain, but this can be mitigated by choosing moderately large enough domain sizes. 

\begin{figure}[htbp]
    \centering
    \subfloat[][$\epsilon = 0.1$]{
    \includegraphics[width=0.48\textwidth]{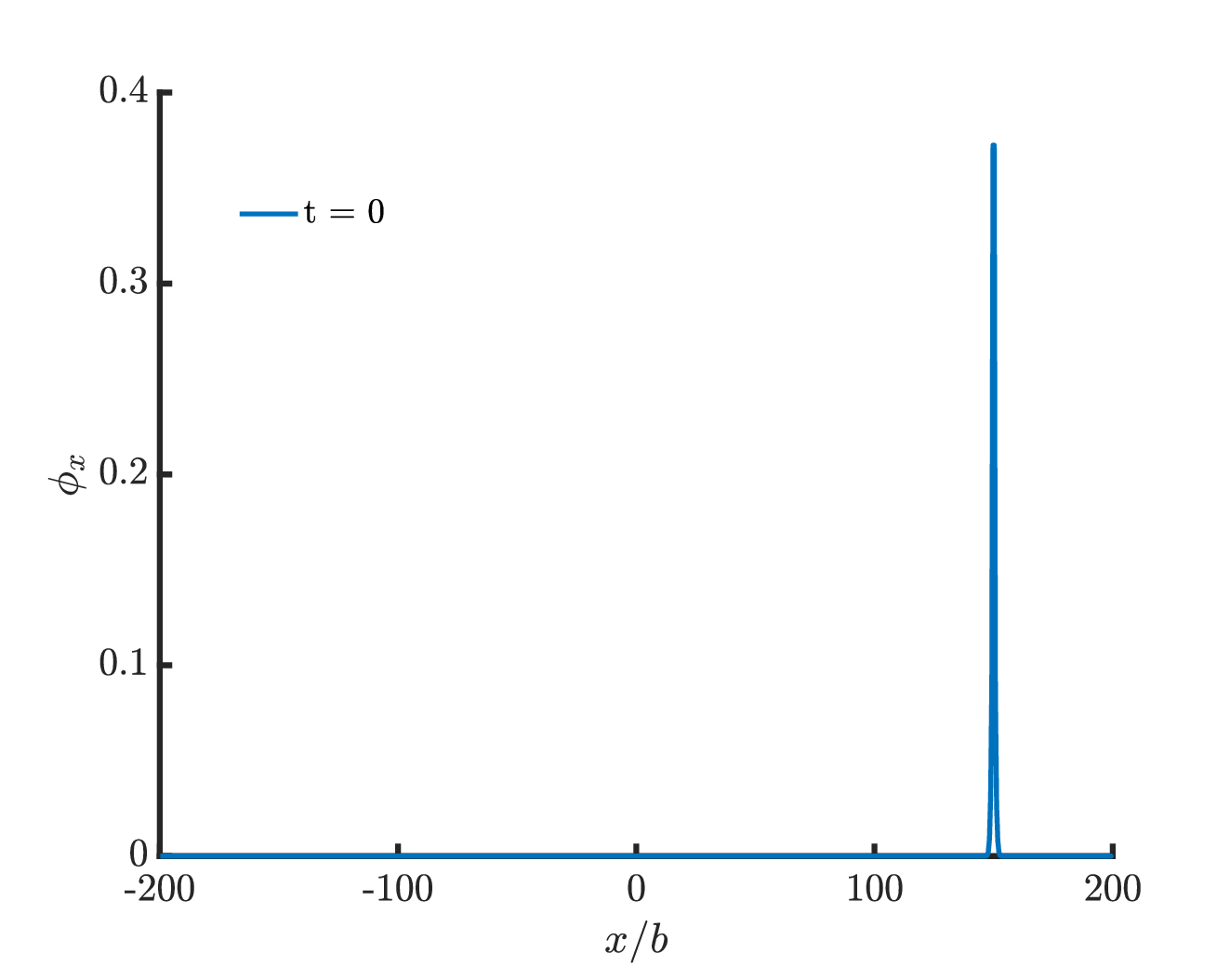}} 
    \subfloat[][$\epsilon = 0.01$]{
    \includegraphics[width=0.48\textwidth]{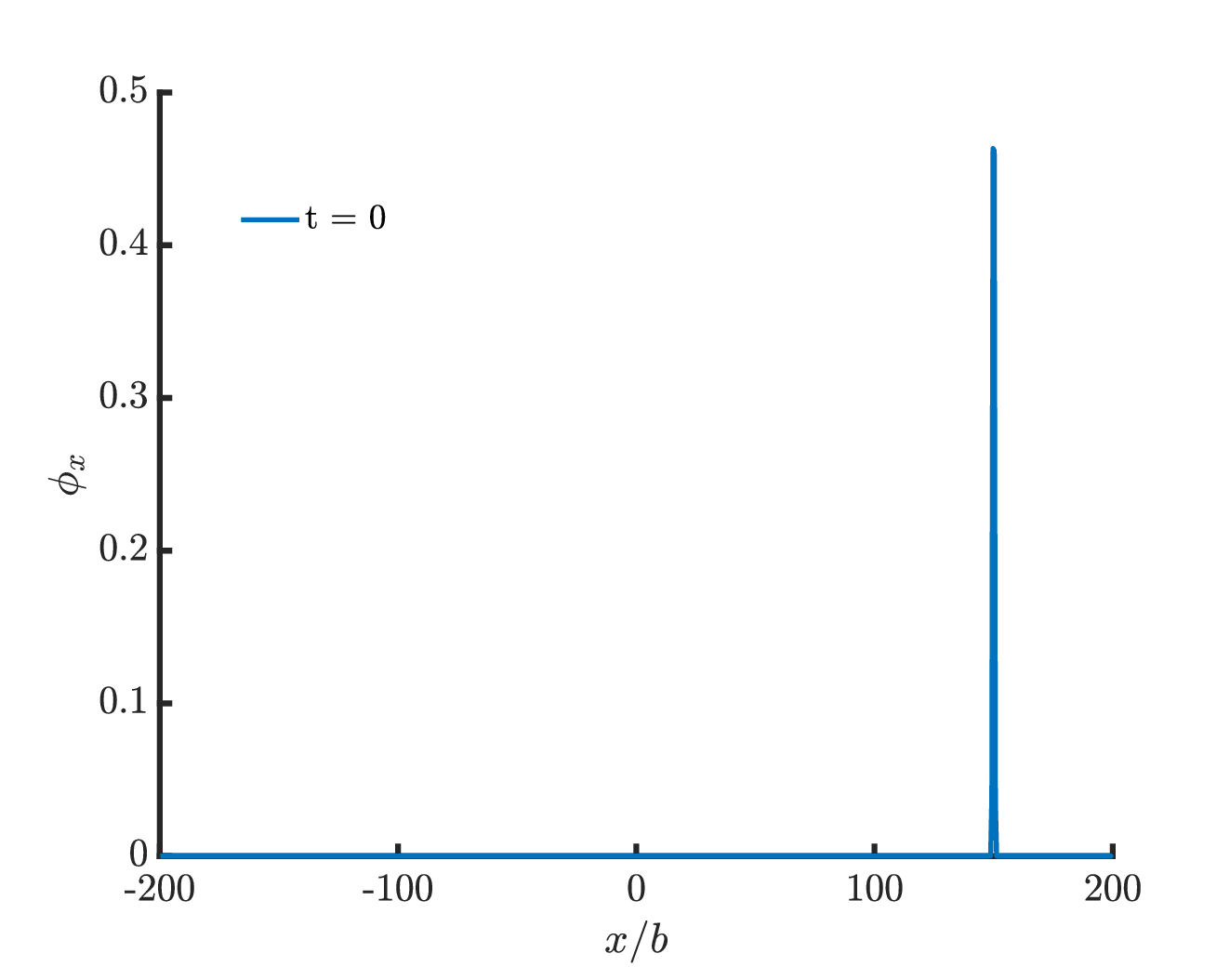}}
    \caption{The equilibrated rupture front $(\phi_x)$ profile at $\sigma_{ap}/ \mu = -0.01$ for different values of $\epsilon$.}
    \label{fig:IC_front_400_400}
\end{figure}

We use quasistatic evolution to study the long-time behavior of the moving rupture front. Firstly, the rupture front is equilibrated with applied normal stress $\sigma_{ap} = -0.01$, and these are shown in Fig.~\ref{fig:IC_front_400_400} for different values of $\epsilon$. Then, the equilibrated front is driven under shear stress loading $\tau_{ap}/\mu = 0.2$, and the long-time behavior is studied. The non-dimensional domain size, fault layer width, and mesh size in the fault layer for the simulations used here are shown in Table \ref{tab:simulation_asymptotic}. The domain size has been increased to probe the long-time behavior of the moving rupture front.

\begin{table}[htbp]
    \centering
    \begin{tabular}{|c|c c c c c|}
    \hline
    \textbf{Length}  & $H/b$ & $W/b$ & $l/b$ & $\Delta h_x/b$ &  $\Delta h_y/b$ \\ 
         \hline
    \textbf{Value} & 400 & 400 & 2 & 0.5 & 0.5  \\
         \hline
    \end{tabular}
    \caption{Domain size, fault-layer width, and mesh size used in simulations for studying the long-time behavior of rupture front.}
    \label{tab:simulation_asymptotic}
\end{table}

The parameters used for simulations here are given in Table \ref{tab:parameters_asymptotic}. We consider the case with parameter $a = 0$, as the speed of the rupture front is large as compared to when $a$ is non-zero, and the long-time behavior of the front can be probed relatively quickly with simulations.

\begin{table}[htbp]
    \centering
    \begin{tabular}{|c|c c c c c c c c|}
    \hline
  \textbf{Parameter}  & $q$ & $a$ & $\epsilon/(\mu b^2)$ & $\lambda/ \mu$ & $B b/\sqrt{\mu \rho}$ & $\tilde{\lambda}$ & $\overline{\phi}$ & $\kappa$\\
         \hline
   \textbf{Value} & 1 & 0 & 0.1 \& 0.01 & 2.25 & 1.0 & 0.7 & 0.5 & 0.695 \\
         \hline
    \end{tabular}
    \caption{Parameters used in simulations for studying the long-time behavior of rupture front.}
    \label{tab:parameters_asymptotic}
\end{table}

\begin{figure}[ht!]
    \centering
    \subfloat[][$\epsilon = 0.1$]{
    \includegraphics[width=0.48\textwidth]{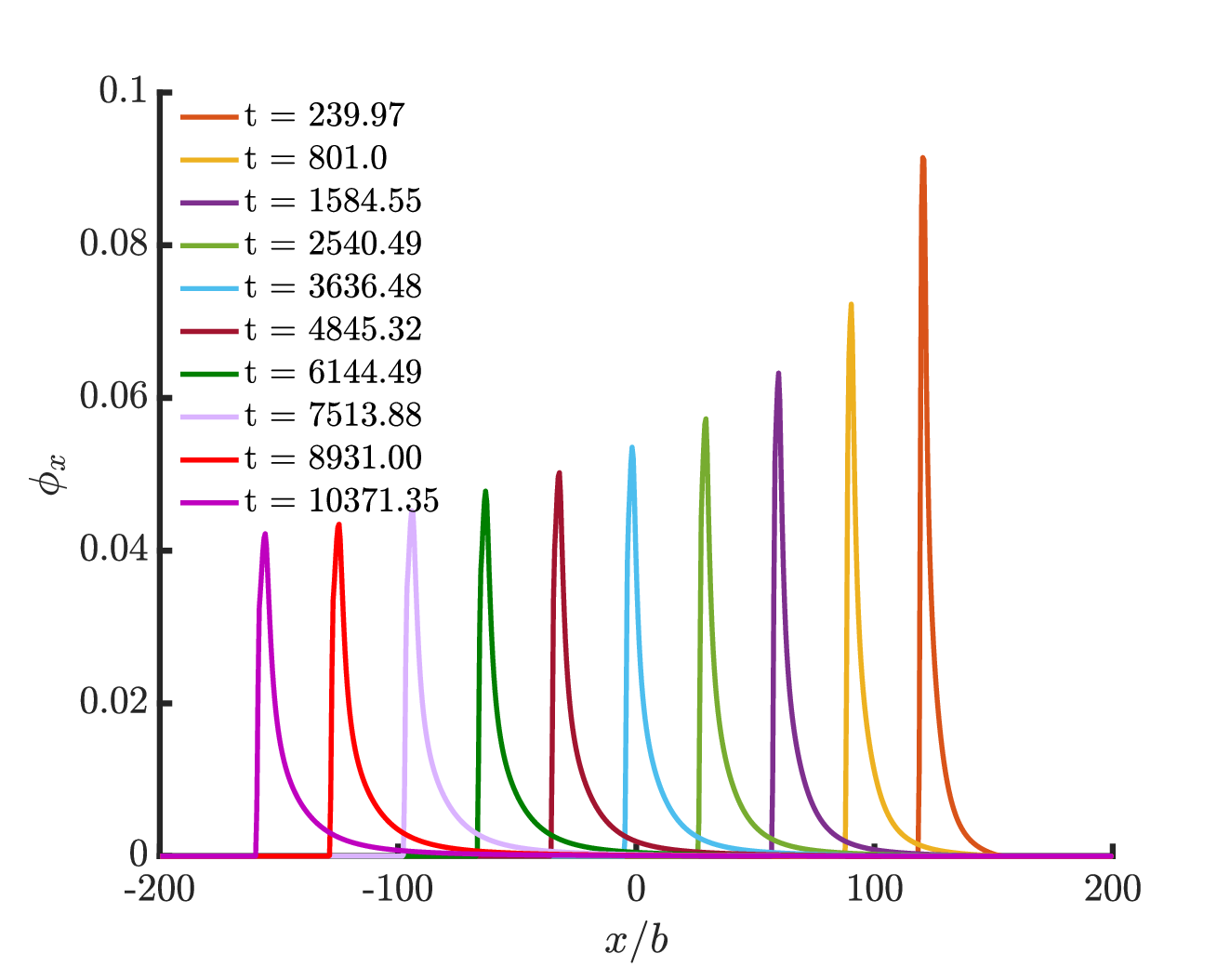}} 
    \subfloat[][$\epsilon = 0.01$]{
    \includegraphics[width=0.48\textwidth]{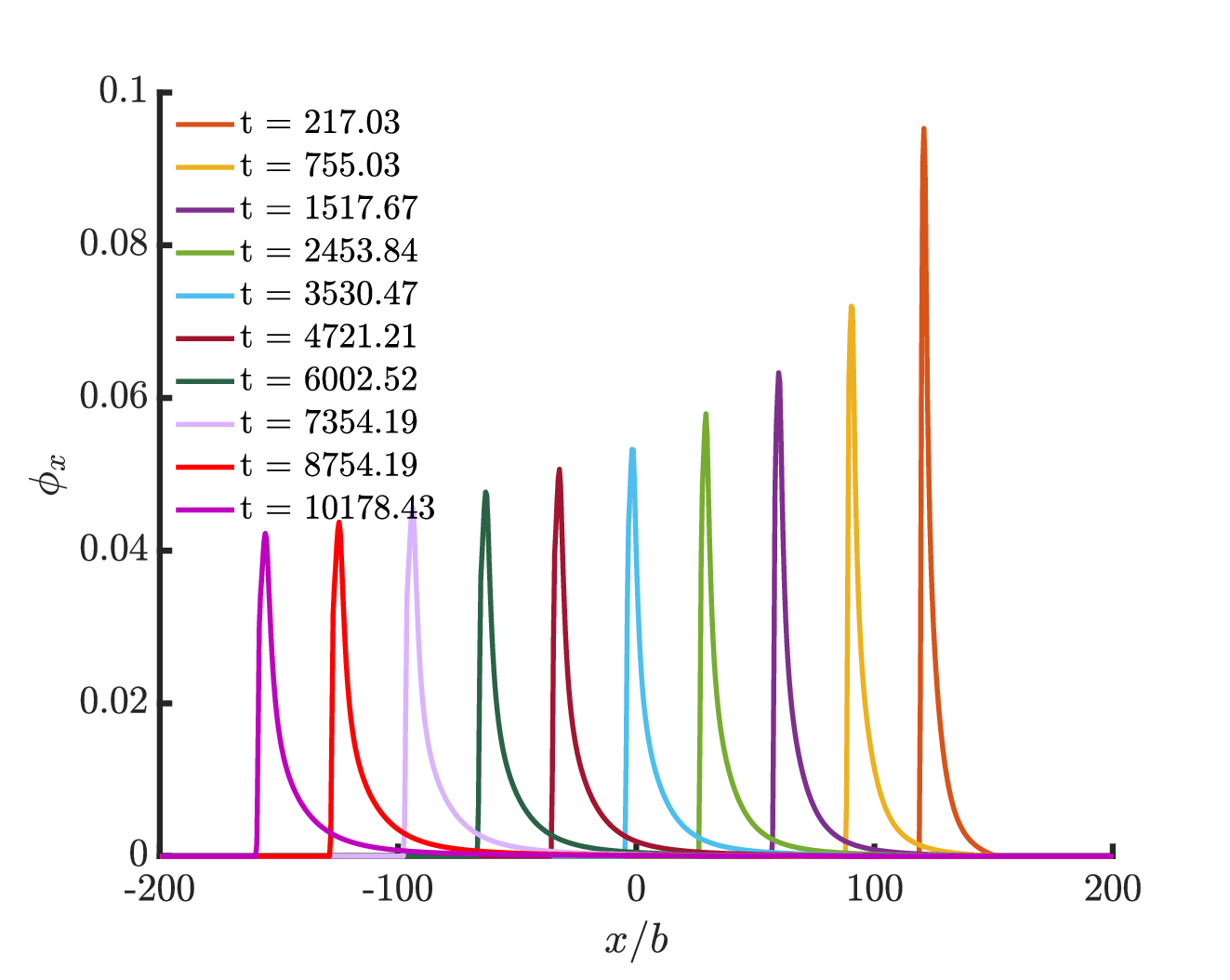}}
    \caption{The propagating rupture front for the crack case under $\tau_{ap}/ \mu = 0.2$ and $\sigma_{ap}/ \mu = -0.01$ for different values of $\epsilon$. It must be noted that the scale of the $y$ axis in this Figure is different than the one used in Fig.~\ref{fig:IC_front_400_400} for plotting the equilibrated profiles.}
    \label{fig:asymptotic_behavior_front_400_400}
\end{figure}

\begin{figure}[ht!]
    \centering
    \includegraphics[scale = 0.42]{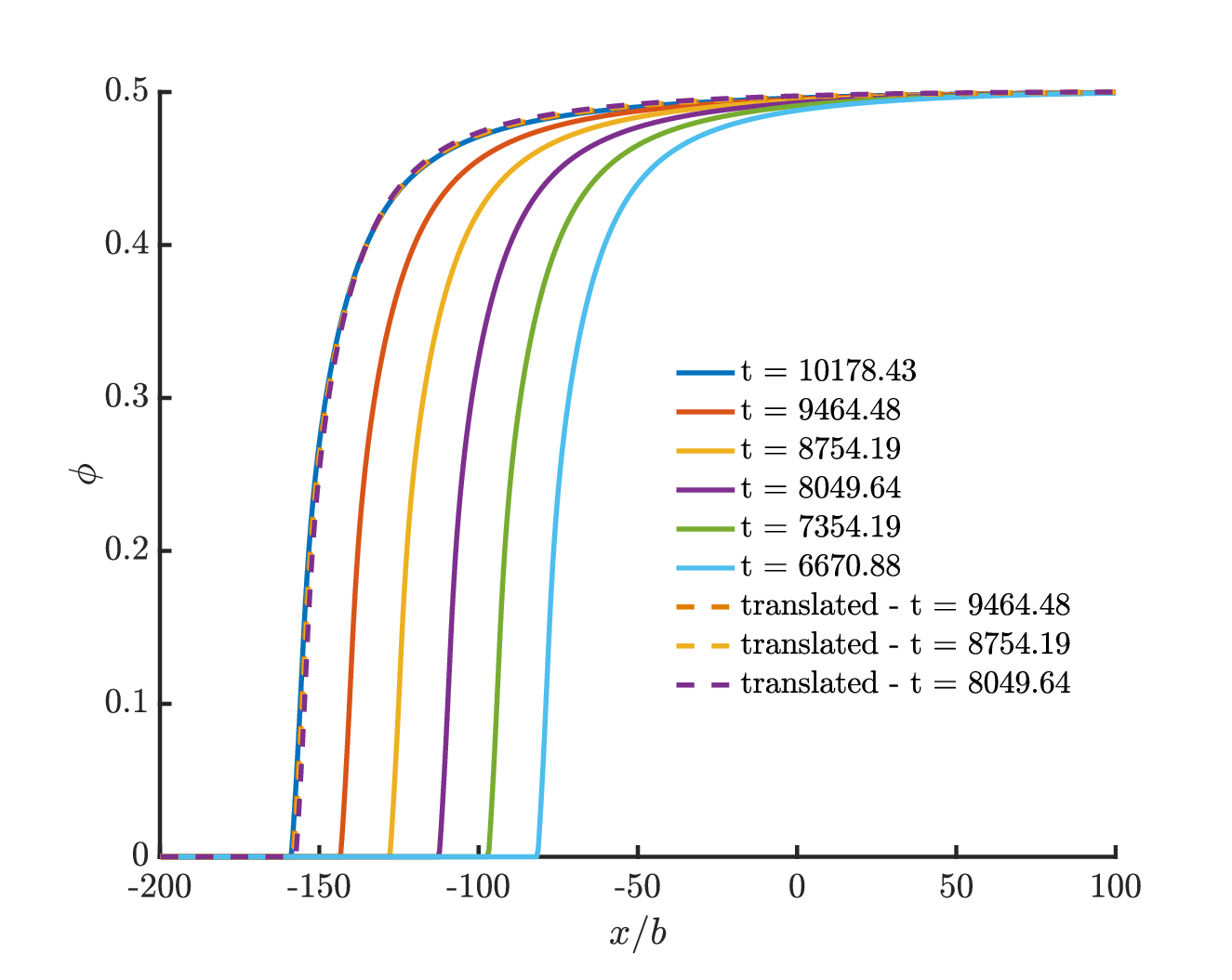}
    \caption{Asymptotic profile of the rupture front $\phi$ for the crack case under $\tau_{ap}/ \mu = 0.2$ and $\sigma_{ap}/ \mu = -0.01$, with  $\epsilon = 0.01$.}
    \label{fig:asymptotic_phi_400}
\end{figure}

As shown in Fig.~\ref{fig:asymptotic_behavior_front_400_400}, the equilibrated profile ($\phi_x$) at $t=0$, shown in Fig.~\ref{fig:IC_front_400_400}, changes its shape while moving under the applied shear stress, for both values of non-dimensional $\epsilon$'s. This suggests that for the governing equations considered in this model, the equilibrated profiles obtained at zero shear stress are not traveling wave profiles when the shear stress is applied. A closer look at the $\phi_x$ profile for large times in Fig.~\ref{fig:asymptotic_behavior_front_400_400} suggests that the front profile becomes more or less stable, and remains localized after some point in time. To illustrate this carefully, we compare the $\phi$ profile for $\epsilon = 0.01$ at $t = 10178.43$, and we translate the $\phi$ profiles at some times to match with the leading edge of $\phi$ profile at $t = 10178.43$. As can be observed in Fig.~\ref{fig:asymptotic_phi_400}, the $\phi$ profiles have become stabilized from $t = 8049.64$ to $t = 10178.43$, and the rupture front remains localized. This suggests that the shear-loaded rupture front evolves for some time and then becomes stabilized, which suggests that it is likely that a limiting traveling wave profile is attained in time.

\subsection{Test of supershear behavior for a crack front under impact loading} \label{sec:supershear}
\begin{figure}[htbp]
    \centering
    \subfloat[][]{
    \includegraphics[width=0.5\textwidth]{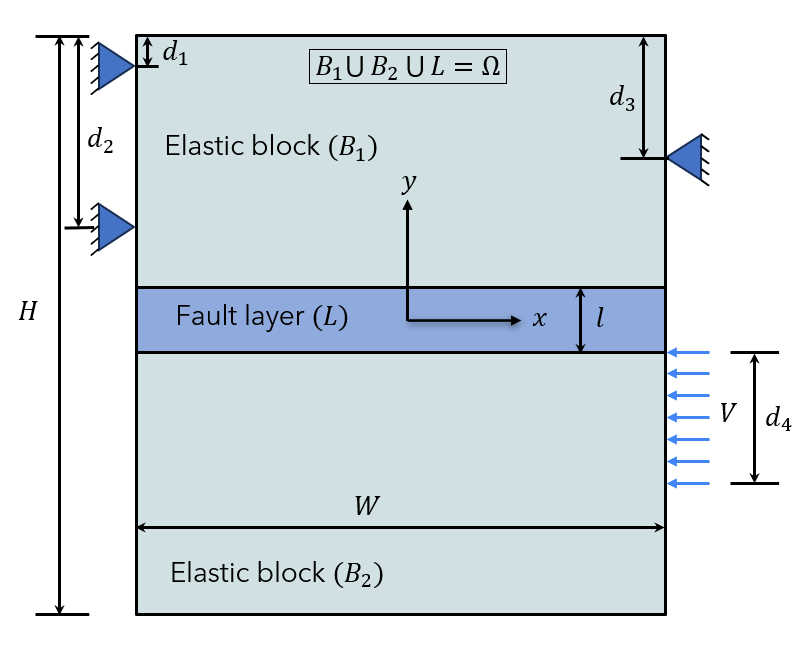}}
    \subfloat[][]{
    \includegraphics[width=0.4\textwidth]{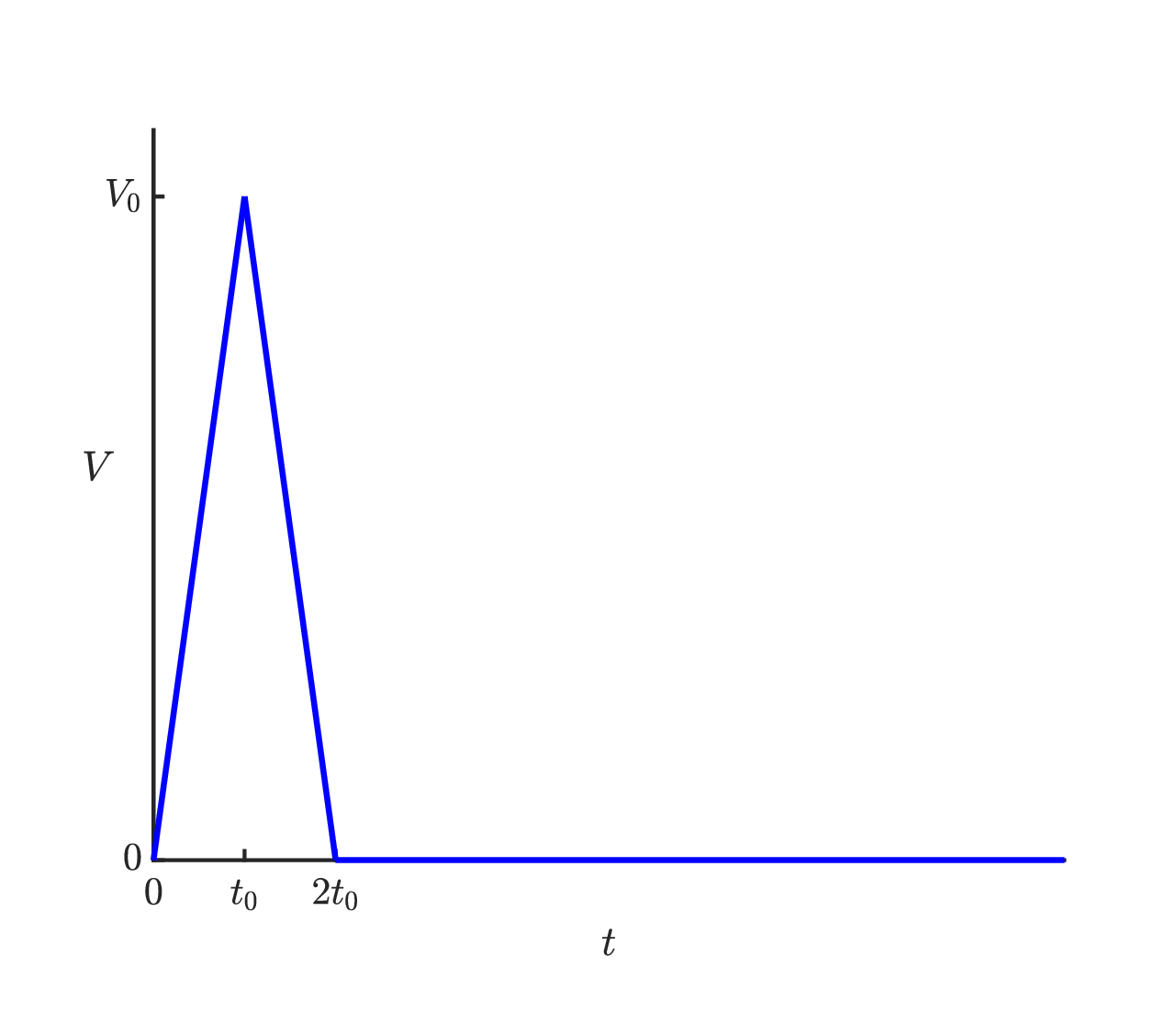}}
    \caption{(a) Schematic of impact loading simulation for a crack (without any pre-strain), (b) An approximate impact velocity boundary condition used in the simulation.}
    \label{fig:schematic_impact}
\end{figure}

We model the impact loading experiments performed on a monolithic cracked body \cite{rosakis1999cracks} using the continuum model developed in this work. A monolithic body corresponds to the case when the elastic modulus of the virgin material in the fault layer is the same as in elastic blocks, meaning $\Tilde{\lambda} = 1.0$. However, as designed in the experiments performed in \cite{rosakis1999cracks}, the crack moves only through a weak bond line, which is modeled here by allowing plastic straining and subsequently damage only in the fault layer. The schematic of the impact loading test for a body with a crack (and without any pre-strain) is shown in Fig.~\ref{fig:schematic_impact}(a), and the impact velocity boundary condition with time is shown in Fig.~\ref{fig:schematic_impact}(b). 

The domain size, the fault layer width, and lengths for impact loading boundary conditions (refer to Fig.~\ref{fig:schematic_impact}), and the mesh size in the fault layer used for the simulations in this Section are shown in Table \ref{tab:simulation_super_shear}.  

\begin{table}[htbp]
    \centering
    \begin{tabular}{|c|c c c c c c c c c|}
    \hline
   \textbf{Length}  & $H/b$ & $W/b$ & $l/b$ & $d_1/b$ & $d_2/b$ & $d_3/b$ & $d_4/b$ & $\Delta h_x/b$ &  $\Delta h_y/b$ \\ 
    \hline 
    \textbf{Value} & 300 & 300 & 2 & 10 & 110 & 60 & 74.5 & 0.5 & 0.5 \\
    \hline
    \end{tabular}
    \caption{Domain size, fault-layer width, lengths for boundary conditions in impact loading, and mesh size in the fault layer, for impact loading simulations of a crack in a body.}
    \label{tab:simulation_super_shear}
\end{table}
 
\begin{table}[htbp]
    \centering
    \begin{tabular}{|c|c c c c c c c c c c|}
    \hline
   \textbf{Parameter} & $q$ & $a$ & $\epsilon/(\mu b^2)$ & $\lambda / \mu$ & $B b/\sqrt{\mu \rho}$ & $\tilde{\lambda}$ & $\kappa$ & $\overline{\phi}$ & $V_0/v_s$ & $(t_0 v_s)/b$\\ 
    \hline
   \textbf{Value} & 1 & 0 & 10 & 2.25 & 0.1534 & 1.0 & 0.995 & 0.5 & 50 & 1.0 \\
    \hline
    \end{tabular}
    \caption{Parameters used for impact loading simulations for a crack.}
    \label{tab:parameters_super_shear}
\end{table}

The parameters taken for the impact loading simulations are shown in Table \ref{tab:parameters_super_shear}. In particular, we work with the $a=0$ case, which corresponds to the case of no relative gain or loss in the strength of the material under compressive loading, as modeled in the damage energy function $(\eta)$ given in \eqref{eta_function}. This is chosen based on the reasoning that for the impact loading case, the specimen shown in Fig.~\ref{fig:schematic_impact}(a) is not put under global compressive loading. Moreover, the impact resistance to motion of the front is significantly less for the $a=0$ case, and importantly, this case is more suitable for the problems explored in this Section. We are primarily interested in finding the impact velocities at which the speed of the rupture/crack front becomes greater than the shear wave speed of the material, which is known as the state of the supershear as observed in the experiments in \cite{rosakis1999cracks}. 

\begin{figure}[ht!]
    \centering
    \subfloat[][t=0]{
    \includegraphics[width=0.49\textwidth]{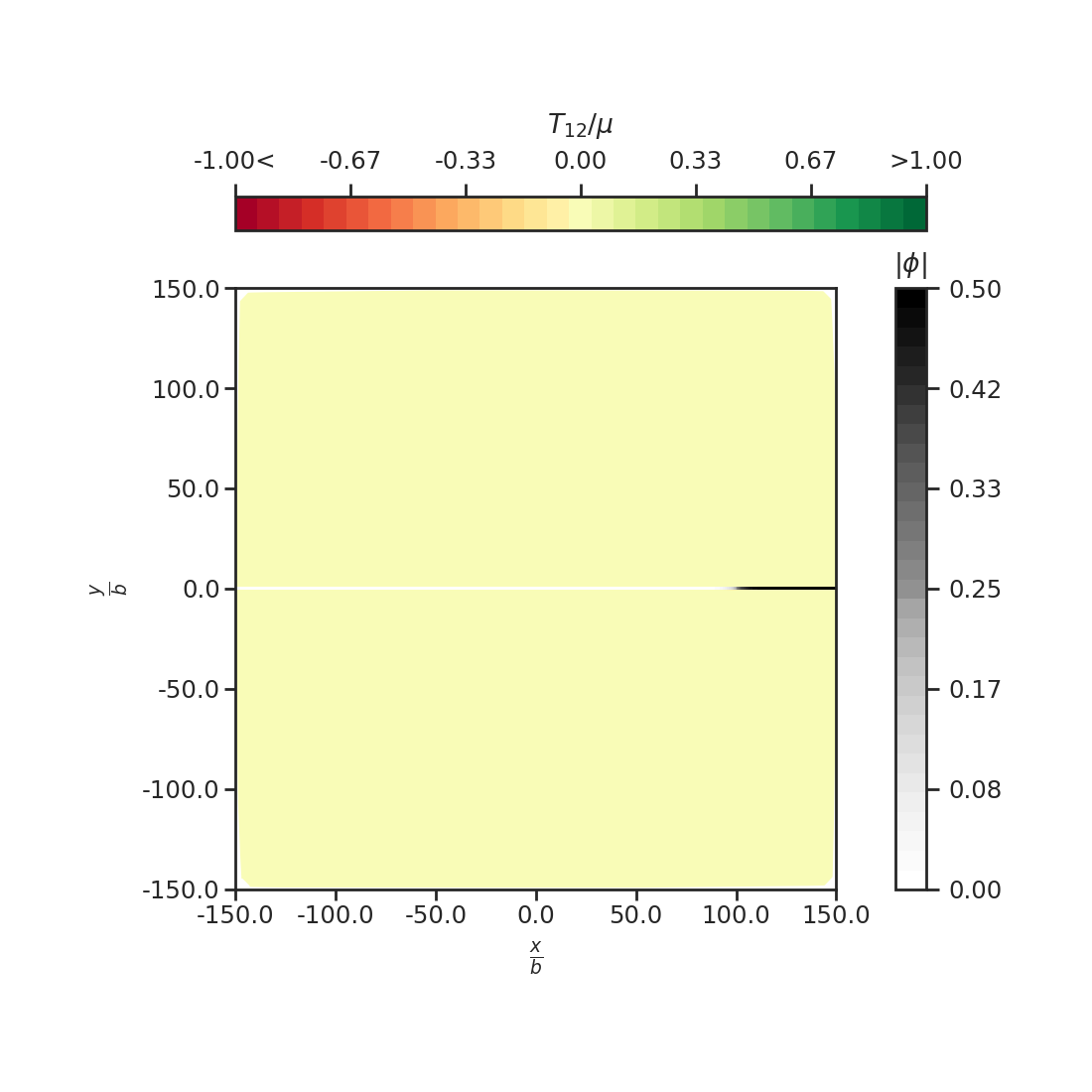}}
    \subfloat[][t=26.84]{
    \includegraphics[width=0.49\textwidth]{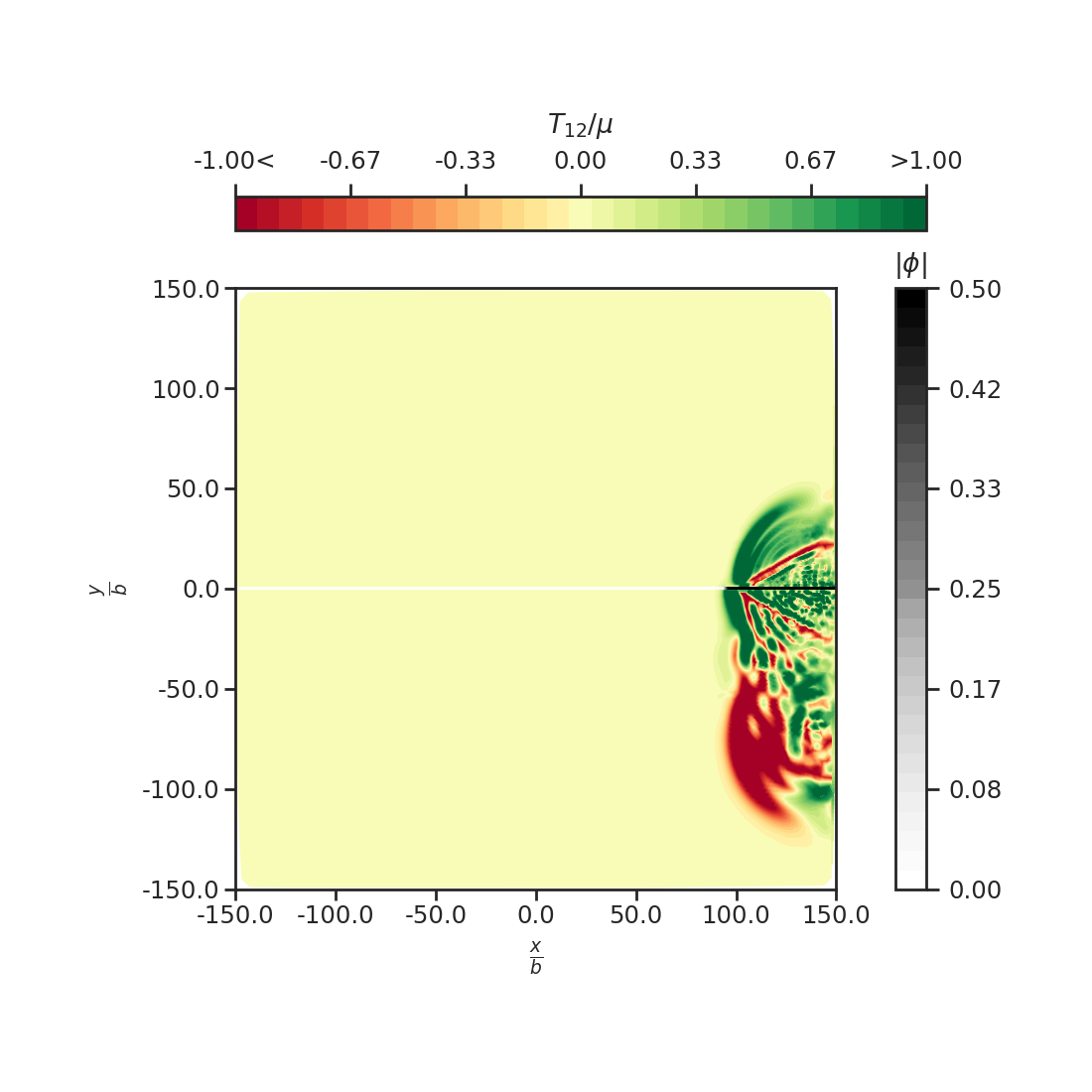}}\\
    \subfloat[][t=45.98]{
    \includegraphics[width=0.49\textwidth]{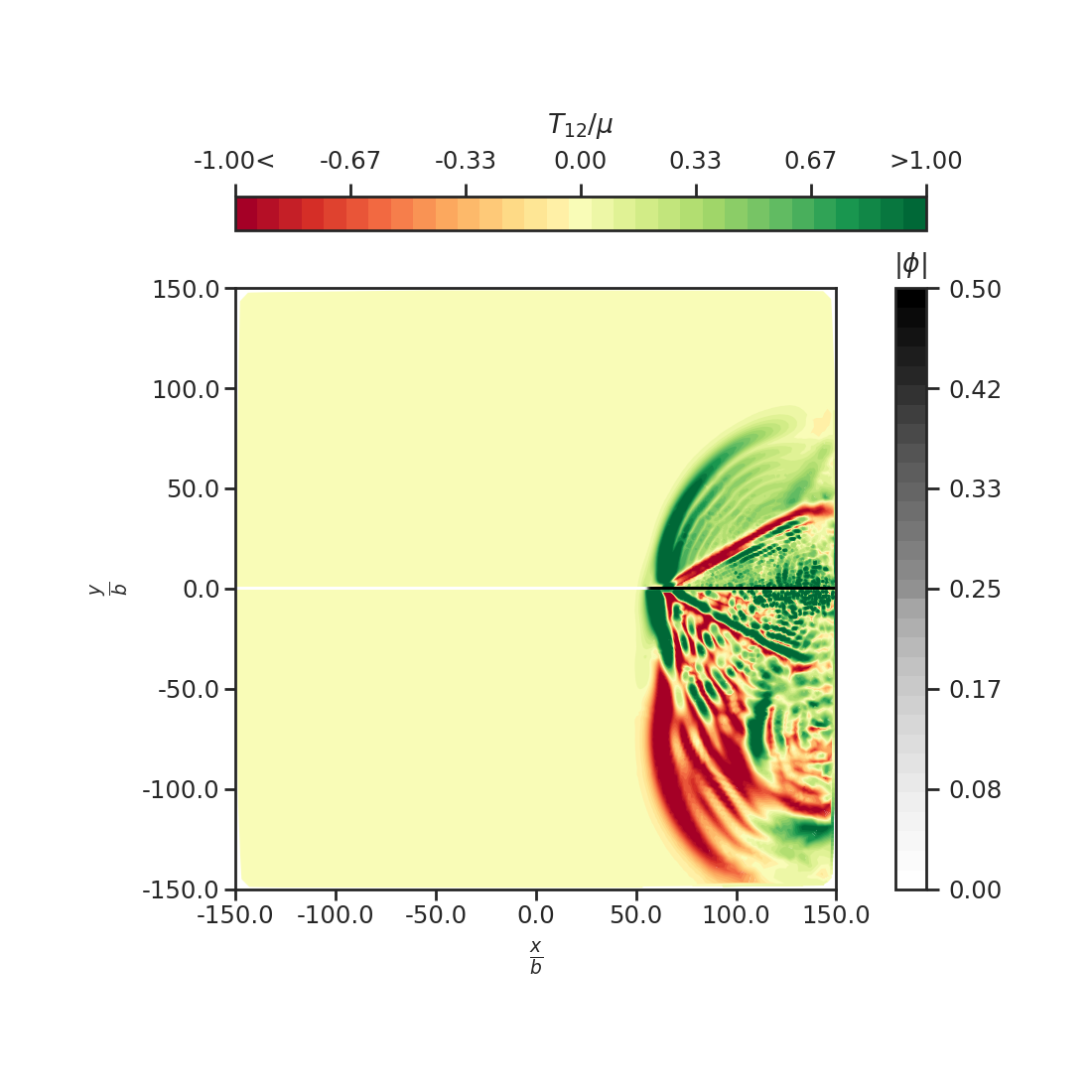}} 
    \subfloat[][t=105.11]{
    \includegraphics[width=0.49\textwidth]{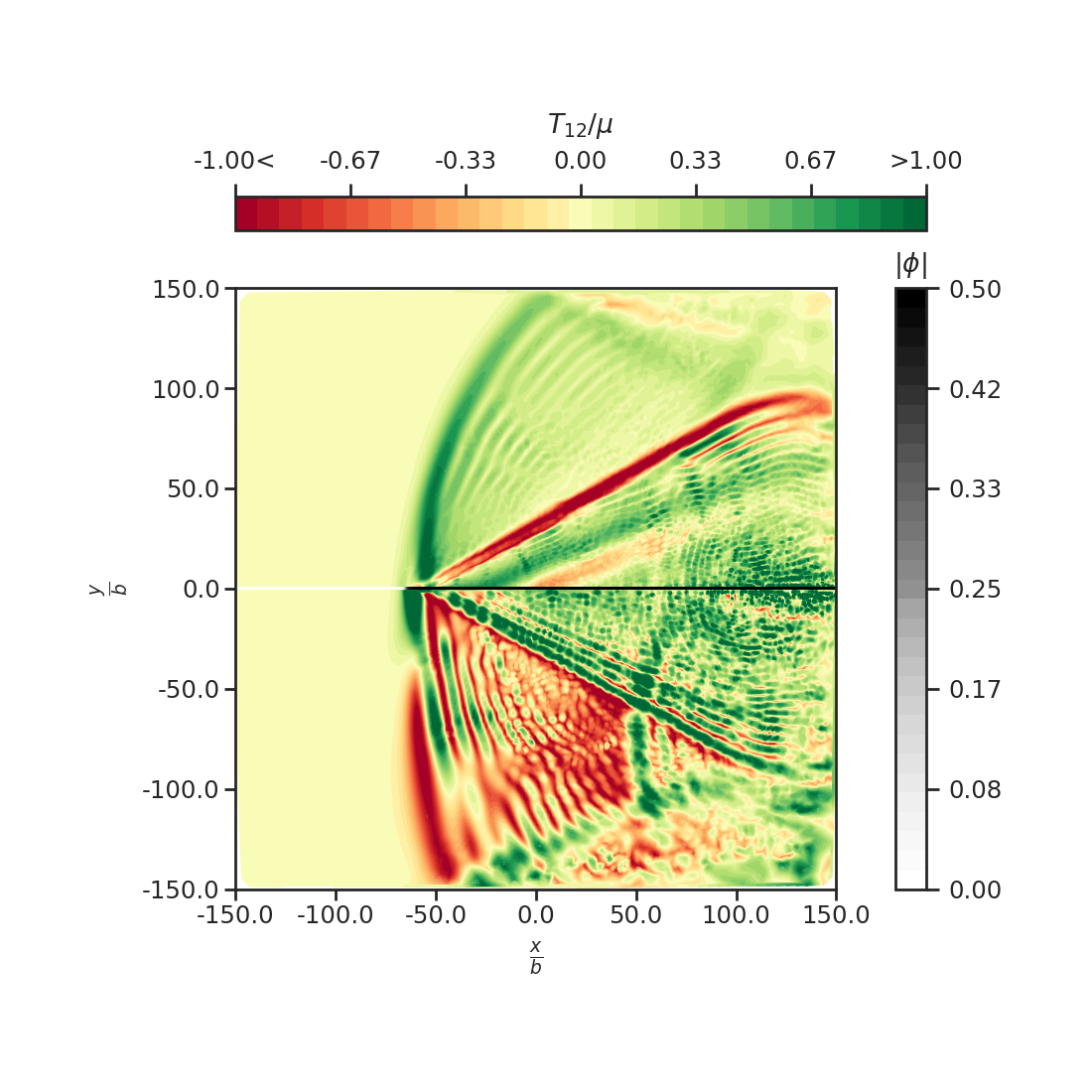}}
    \caption{The propagating rupture/crack front and the shear wave ($T_{12}/ \mu$) under impact loading and without any pre-strain, at different non-dimensional times.}
    \label{fig:impact_load_shear_wave}
\end{figure}

\begin{figure}[ht!]
    \centering
    \subfloat[][t=0]{
    \includegraphics[width=0.49\textwidth]{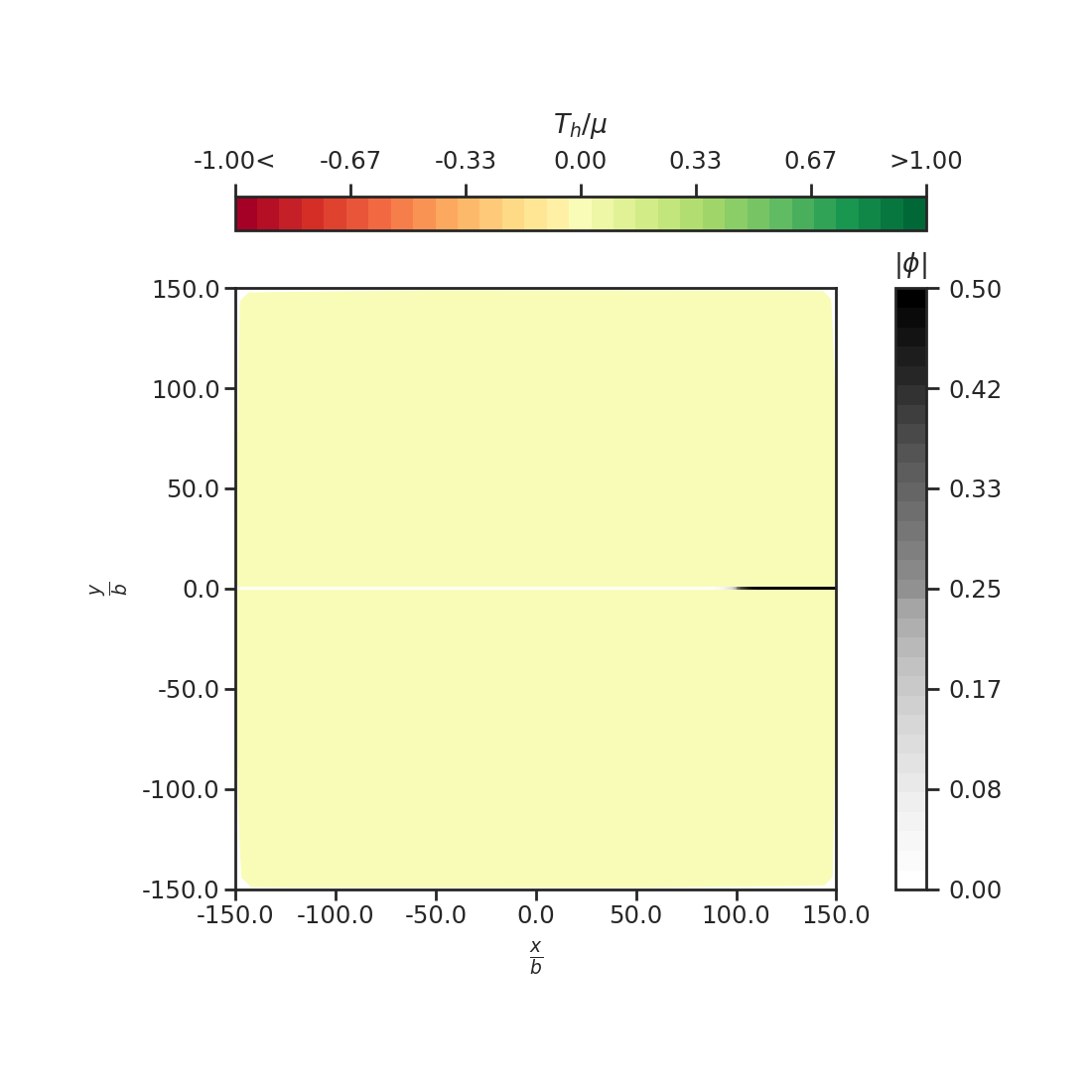}}
    \subfloat[][t=26.84]{
    \includegraphics[width=0.49\textwidth]{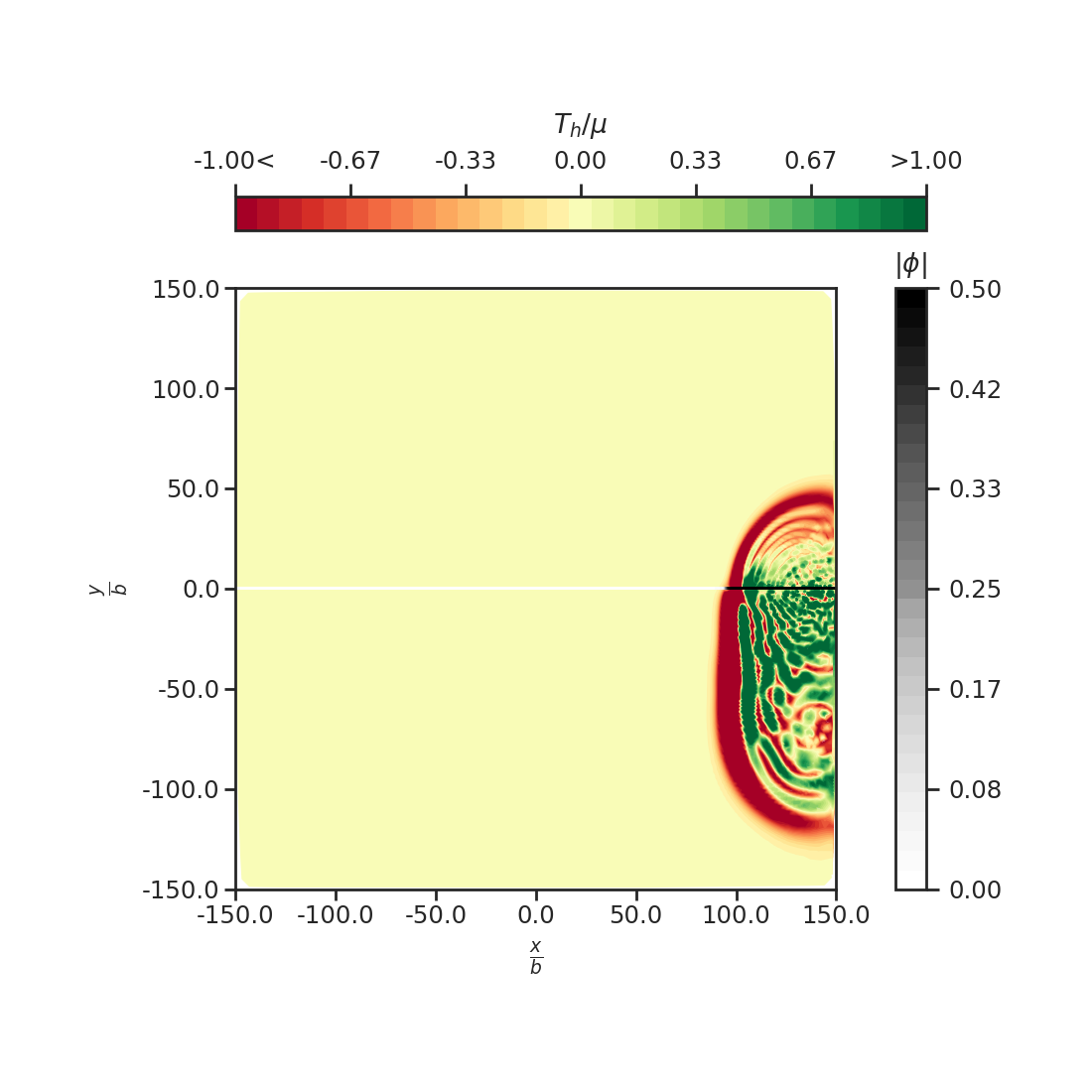}} \\
    \subfloat[][t=45.98]{
    \includegraphics[width=0.49\textwidth]{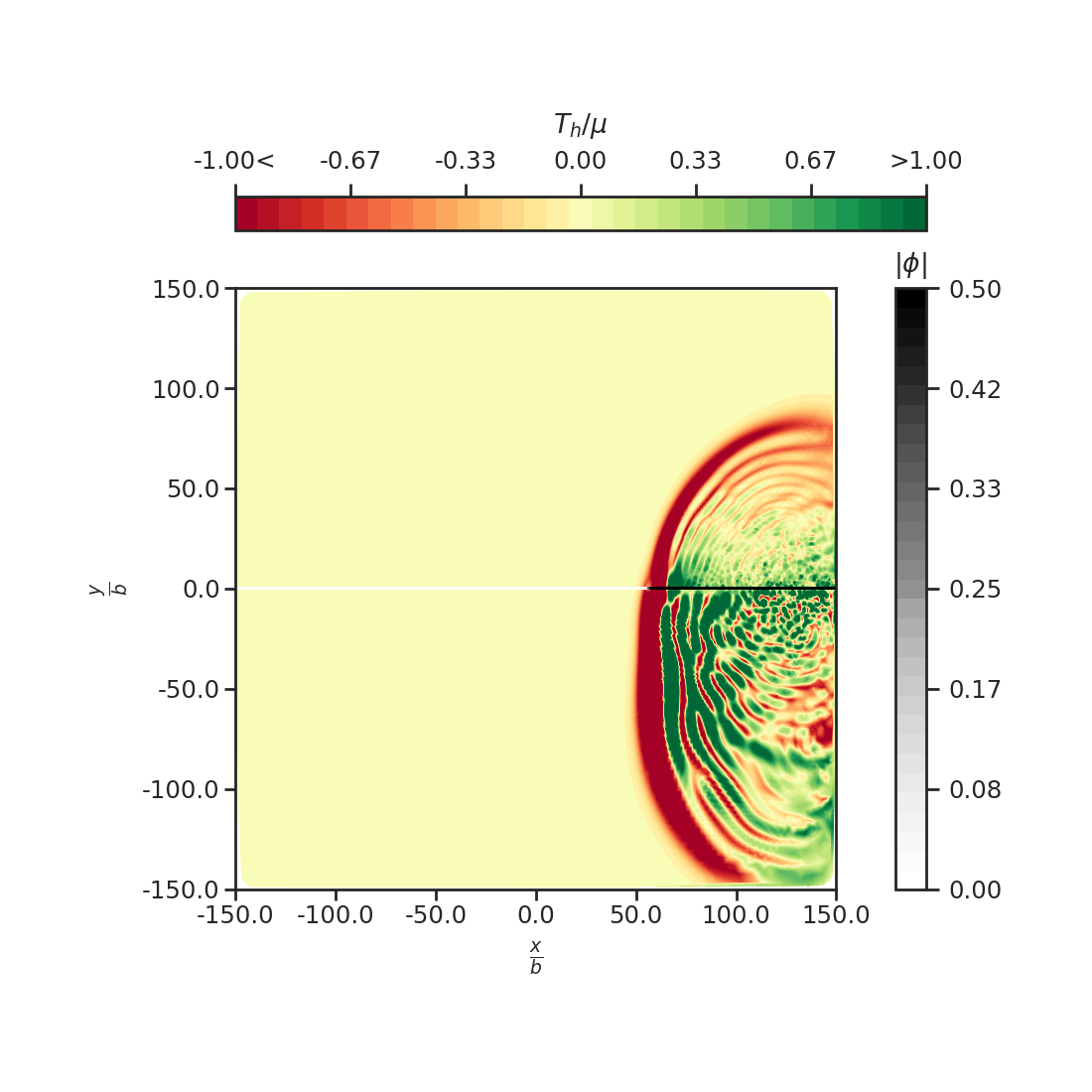}} 
    \subfloat[][t=105.11]{
    \includegraphics[width=0.49\textwidth]{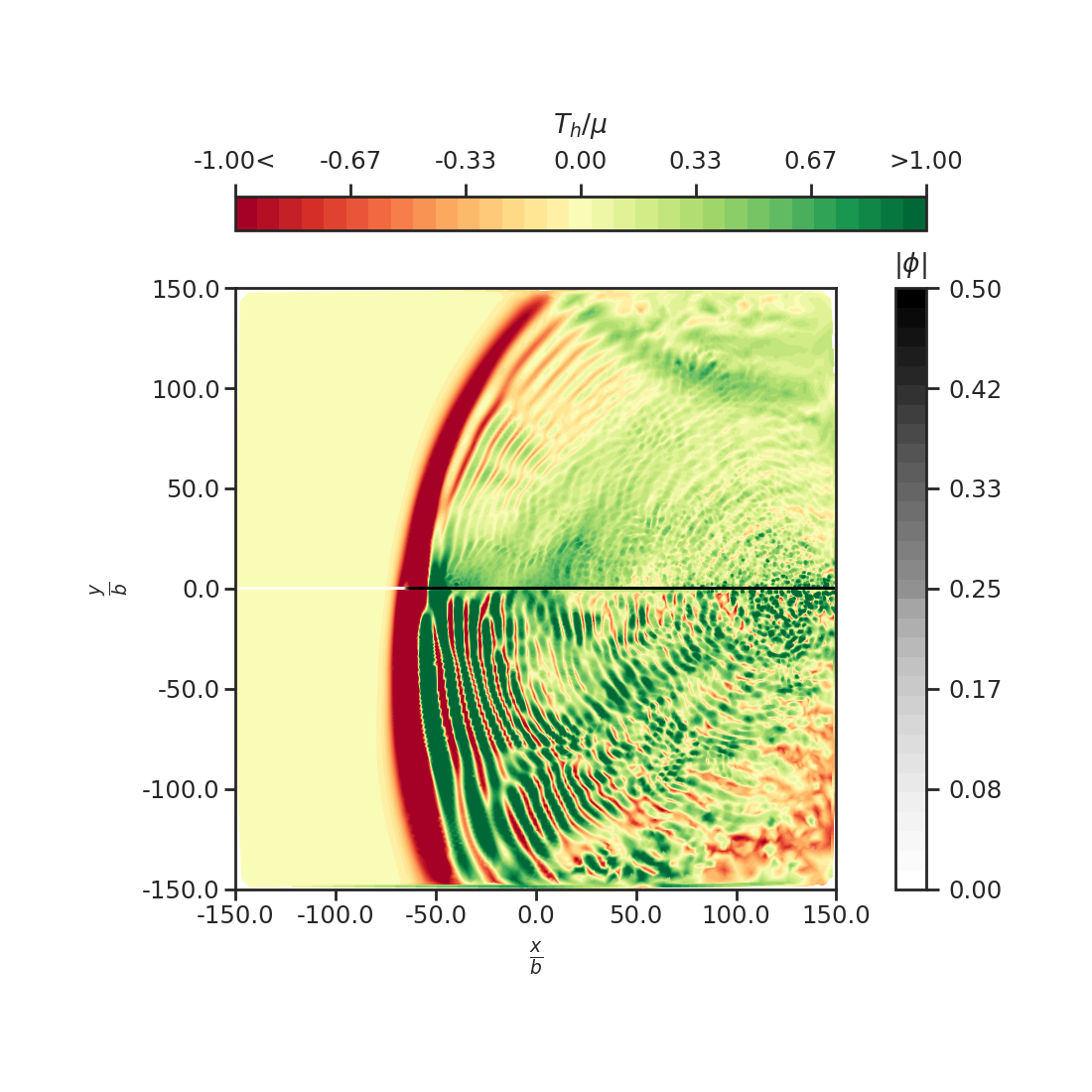}}
    \caption{The propagating rupture/crack front and the pressure wave ($T_{h}/ \mu$) under impact loading and without any pre-strain, at different non-dimensional times.}
    \label{fig:impact_load_pressue_wave}
\end{figure}

We choose the value of $\Tilde{B}=0.1534$, using the same reasoning as in \cite{morin2021analysis}. A dimensionally consistent estimate of crack/rupture front velocity (dimensional) based on local material parameters is
\begin{equation}
\begin{aligned}
    v_c  \approx \frac{1}{B} \frac{q \mu}{ \sqrt{ \epsilon / (q \mu)}} 
      =  \frac{1}{B} \frac{q \mu}{ \sqrt{  \frac{\Tilde{\epsilon} \mu b^2 }{ (q \mu)}}}.
\end{aligned}  
\end{equation}
The ratio of the crack speed to the dilatational wave speed of the material is
\begin{equation}
    \frac{v_c}{v_p} = \frac{v_c}{v_s} \frac{1}{\sqrt{ \frac{\lambda}{\mu} + 2}}.
\end{equation}
Substituting for the estimate of $v_c$ in the above expression we get
\begin{equation} \label{eq:vc_vp}
    \frac{v_c}{v_p} \approx \frac{1}{\Tilde{B}} \frac{q}{ \sqrt{\frac{\Tilde{\epsilon}}{q}} \sqrt{ \frac{\lambda}{\mu} + 2} }.
\end{equation}
Now, the value of $\Tilde{B}$ is chosen such that the crack velocity $v_c$ is the same as the dilatational wave velocity $v_p$. Doing this calculation, for the parameters used for the impact loading simulation (other than $\Tilde{B}$), we get the value of $\Tilde{B} = 0.1534$.

Fig.~\ref{fig:impact_load_shear_wave} and \ref{fig:impact_load_pressue_wave} show the shear ($T_{12}$) wave and the pressure ($T_h$) wave for the impact loading simulation of a crack in a monolithic body (without any pre-strain), and with the parameters taken as shown in Table \ref{tab:parameters_super_shear}. It has been observed in our simulations that the average crack front velocity $(v_c)_{obs}\approx 2.04 \: v_s$, which is in the intersonic regime, and we see the formation of Mach cone in the shear stress wave shown in Fig.~\ref{fig:impact_load_shear_wave}. The crack front speed is less than the dilatational wave speed (for the parameters taken here), and hence, we do not observe the formation of the Mach cone in the pressure wave in our simulations as shown in Fig.~\ref{fig:impact_load_pressue_wave}. The maximum speed at which the crack front moves is the dilatational wave speed, even when the impact velocity is as high as $V_0/v_s = 500$. A similar result is obtained with non-dimensional drag taken as $B=0$ in our simulations, which corresponds to the asymptotic estimate of $v_c \to \infty$ in \eqref{eq:vc_vp} representing an infinitely fast time scale of crack evolution (even in comparison to the time scale of elastic wave propagation). 

To simulate the case with $B=0$, the quasistatic version of $\phi$ equation is solved by introducing a fictitious time scale. The real-time scale of the impact simulation (with $B=0$) is governed just by the stress waves propagation, and for each time step, the $\phi$ equation is iteratively solved using a Picard's iteration, until a desired tolerance on the norm of $\dot{\phi}$ is reached. With this setup, the system of equations is evolved in physical time.

To summarize, we have computationally demonstrated supershear behavior of the crack front under impact loading as obtained in the experiments in \cite{rosakis1999cracks}. However, our results also indicate that it is not possible to obtain supersonic motion of the crack front beyond the dilatational wave speed of the bulk material using our model when the boundary loading is solely an initial impact - this is physically correct. The supersonic motion of the crack front can be obtained under pre-strained conditions of a cracked body, as also shown previously in \cite{morin2021analysis, zhang2015single} - this feature is physically plausible, and in accord with molecular dynamics simulations of metals, as discussed in \cite{zhang2015single}.

\section{Conclusion} \label{conclusion}  
In this work, we have developed a mathematical model for studying fault friction in dynamic rupture. We can obtain emergent static fault friction laws from our simulations, without putting in (by hand) any such `switching' criterion in our model. We also make a quantitative prediction of the material constants for geomaterials like cohesion and friction angle, by doing a linear fit (which resembles a Mohr-Coulomb failure curve) to our simulation data points. The physical phenomena of short-slip, slip-weakening, and self-healing are probed in dynamic simulations using our continuum model. We can predict different characteristics in the slip and average shear stress profile in the fault layer, depending upon the amount of elastic damage allowed in the wake of the rupture front. Moreover, the long-time behavior of the moving rupture front under shear stress suggests that the front remains localized. However, the equilibrated profile under no applied shear stress is not a traveling wave profile under applied shear, but potentially transforms into one in the limit of long times and sufficient applied stress. In another study, a crack/rupture front is driven under impact loading without any pre-stress, and it is observed in our simulations that an upper bound for the crack front speed is the same as the dilatational or pressure wave speed of the material, as physically expected. The supershear phenomenon is observed in our results, as observed in the experimental work in \cite{rosakis1999cracks}. Supersonic crack motion can be obtained in our model when the crack/rupture front is present under pre-stressed conditions, as also shown previously in \cite{zhang2015single, morin2021analysis} for a simpler mechanical model (not adapted for geomechanics). 

It will be interesting to work with a model where the rupture front can move in curved fault layers in 2-d instead of in a flat fault layer as in this work. In this case, the scalar H-J equation becomes a multi-D system of H-J equations. However, solving a system of H-J equations is a very challenging task, as no rigorous theory or computational schemes for such equations exist in the mathematical literature, to our knowledge. A dual variational principle can be developed for such difficult PDE and used as a basis for computation, based on the work in \cite{acharya2023dual, acharya2022variational, kouskiya2023hidden, arorathesis2023, singh2024hidden, kouskiya2024hidden}.

It will be instructive to probe the asymptotic behavior of the moving front more rigorously using dynamic evolution, which requires the inclusion of perfectly matched layers at the boundaries of the domain to mitigate effects of stress-wave reflection. 

A natural extension of this planar setting is in 3-d, where a planar 2-d fault layer is surrounded by 3-d elastic blocks, and the scalar 1-d H-J equation becomes a scalar multi-d equation, which can be solved relatively easily by extending the numerical schemes used in this work, as demonstrated in \cite{morin2021analysis}. 

\section*{Acknowledgments}
This work was supported by the grant NSF OIA-DMR \#2021019. It is a pleasure to acknowledge helpful discussions with Jacobo Bielak on this work.

\bibliographystyle{alpha}\bibliography{paper_template.bib}

\end{document}